\documentclass[final,3p,authoryear]{elsarticle}

\usepackage{amsmath}
\usepackage{amssymb}
\usepackage{amsthm}

\usepackage{mathtools}
\usepackage{graphicx}

\newtheorem{theorem}{Theorem}

\usepackage{xcolor}

\DeclareMathOperator{\Real}{Re}

\usepackage[font={small,stretch=0.95},figurename={Fig.},labelfont=bf,labelsep=period]{caption}

\usepackage[colorlinks=true]{hyperref}

\begin{document}

\begin{frontmatter}



\title{The topological dynamics of continuum lattice grid structures}

\author[aff1]{Yimeng Sun}

\author[aff1]{Jiacheng Xing}

\author[aff2]{Li-Hua Shao}

\author[aff1]{Jianxiang Wang\corref{cor}}
\ead{jxwang@pku.edu.cn}

\cortext[cor]{Corresponding author.}

\affiliation[aff1]{organization={Department of Mechanics and Engineering Science, College of Engineering, Peking University},
	city={Beijing},
	postcode={100871},
	country={China}}

\affiliation[aff2]{organization={School of Aeronautic Science and Engineering, Beihang University},
	city={Beijing},
	postcode={100191},
	country={China}}

\begin{abstract}

Continuum lattice grid structures which consist of joined elastic beams subject to flexural deformations are ubiquitous. In this work, we establish a theoretical framework of the topological dynamics of continuum lattice grid structures, and discover the topological edge and corner modes in these structures. We rigorously identify the infinitely many topological edge states within the bandgaps via a theorem, with a clear criterion for the infinite number of topological phase transitions. Then, we obtain analytical expressions for the topological phases of bulk bands, and propose a topological index related to the topological phases that determines the existence of the edge states. The theoretical approach is directly applicable to a broad range of continuum lattice grid structures including bridge-like frames, square frames, kagome frames, continuous beams on elastic springs. The frequencies of the topological modes are precisely obtained, applicable to all the bands from low- to high-frequencies.  Continuum lattice grid structures serve as excellent platforms for exploring various kinds of topological phases and demonstrating the topological modes at multiple frequencies on demand. Their topological dynamics has significant implications in safety assessment, structural health monitoring, and energy harvesting.
\end{abstract}




\end{frontmatter}
\section{Introduction}
\label{sec1}
Continuum lattice grid structures that consist of joined uniform elastic beams are pervasive in nature and engineering. They are frames of large-scale structures such as buildings and bridges, skeletons of organisms \citep{Thompson,gibson,SOLON20091331,doi:10.1073/pnas.1705492114, jiangetalpnas2023}, as well as small-scale advanced micro- and nano-structured materials \citep{ashby_properties_2006,schaedler_ultralight_2011,fernandes_mechanically_2021,Xiao2022,yanetalsa2023}. Their dynamics is highly relevant to multifunctionality in wave and vibration control, safety and health monitoring of structures \citep{phani_wave_2006,advs.201900401}, and even to the tissue-modeling process in biology \citep{SOLON20091331,doi:10.1073/pnas.1705492114}.

The topological dynamics of mechanical systems has attracted much attention because of the unique topological modes and immunity to defects \citep{chen2018,Ma2019,XIN2020}. The topological dynamic properties of continuum systems are intriguing due to many more possibilities of topological phase transitions in continuum structures \citep{continuum-PA}, compared to the extensively investigated discrete systems based on the spring--mass model \citep{chen2018,Chan2018,chen2019,wang2023}. Previous topological continuum structures are mostly designed based on the tight-binding model, and characterized by modulations of stiffness and mass, and a few topological phase transitions in specific bandgaps are revealed by numerical calculations. \citet{DUAN2023105251} investigated the edge and corner modes in one bandgap of
a structure composed of weakly interacted plates connected by thin beams using numerical simulations. Other work on topological continuum structures is also based on periodically stiffness- or inertia-modulated structures and numerical computations~\citep{yin_band_2018,fan2019,MUHAMMAD2019359,zhang2019,Zhou2020,practical-SSH}.  The topological dynamics of the most common continuum lattice grid structures with uniform cross-sections and stiffness has not been unveiled. It has been recognized that the topological dynamics of these ubiquitous structures cannot be captured by the tight-binding model, and thus, a theoretical framework for precisely determining the edge states and identifying the topological phase transitions is lacking.

In this work, we discover the topological properties of a broad class of continuum lattice grid structures. Starting with a continuum beam model, we rigorously prove the existence of infinitely many topological states within multiple bandgaps across the whole spectrum by a theorem, and we give a criterion identifying the topological phase transitions. We obtain the analytical expressions of the Zak phase for each frequency band, and then we introduce a topological index determining the edge states (instead of the interface states), which is related to the topological phases of bulk bands and a phenomenon of ``eigenvalue crossing''. The theoretical framework for the continuous beam is directly applied to reveal and demonstrate the ample topological states in a broad range of continuum lattice grid structures including bridge-like frames, square frames and kagome frames, plates, and continuous beams on elastic foundations, torsional springs and linear springs. The frequencies of the edge and corner states in these continuum lattice grid structures are precisely obtained, applicable to any band from low- to high-frequency.


\section{Topological continuous beam model on supports}
We begin with a continuous beam model on supports, as shown in Fig.~1(a). The beam has a uniform cross-section, and it is supported on $2N$ simple supports such that the beam has $2N+1$ segments with alternating lengths $l_2$, $l_1$, $l_2$, \dots, $l_1$, $l_2$. The two ends are clamped.
\begin{figure}[tb!]
	\centering
	\includegraphics[width=0.8\linewidth]{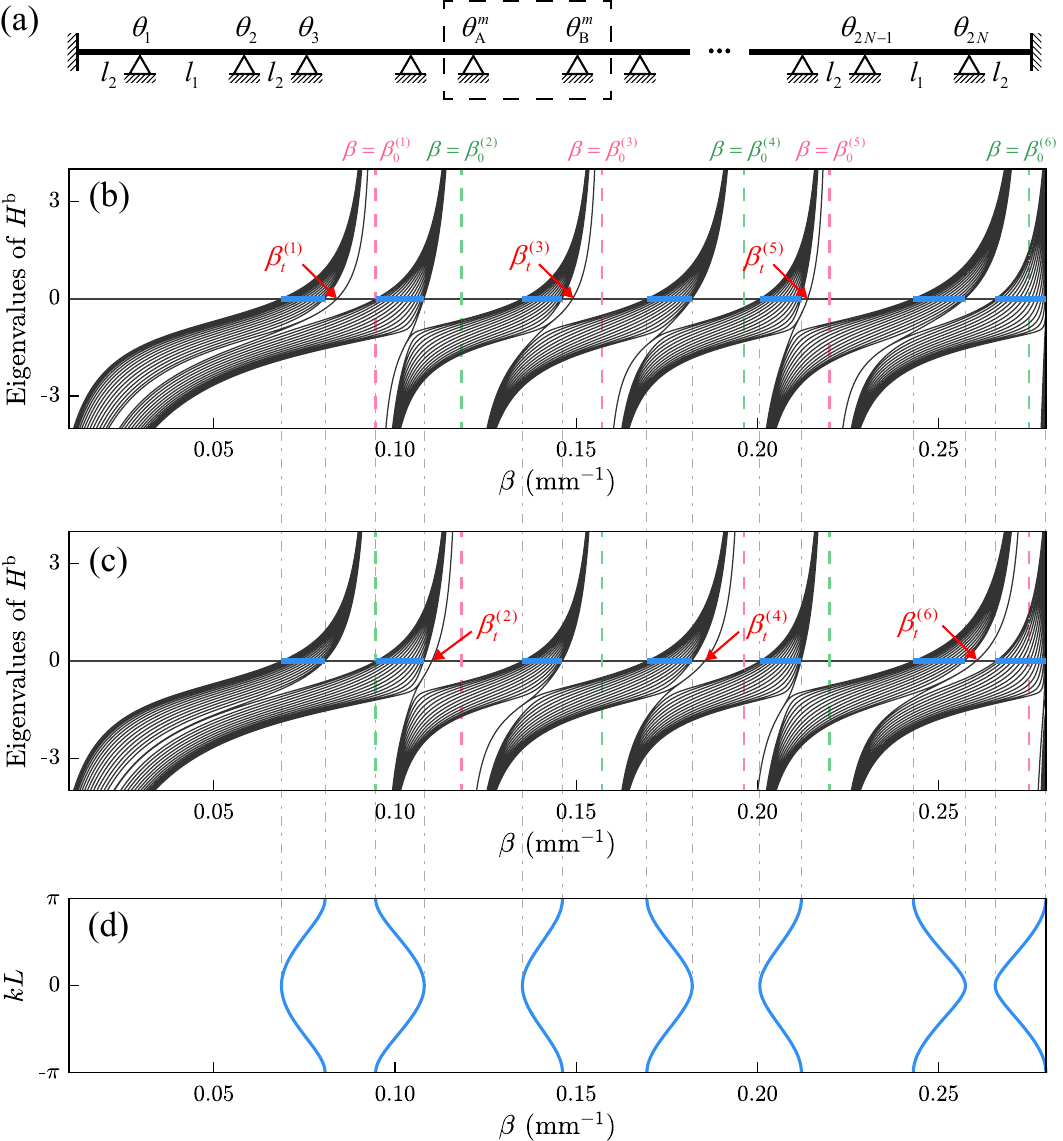}
	\caption{(a) A continuous beam on discrete supports where a unit cell is shown in dashed frame. (b),(c) Spectra of the finite continuous beam with $l_1=40~\mathrm{mm}$, $l_2=50~\mathrm{mm}$ (b), and $l_1=50~\mathrm{mm}$ and $l_2=40~\mathrm{mm}$ (c). Bulk bands are indicated by blue line segments, and topological edge states indicated by red arrows. (d) Dispersion diagram of the  beam  with $(l_1, l_2)=(40~\mathrm{mm},50~\mathrm{mm})$ and $(50~\mathrm{mm},40~\mathrm{mm})$.}
\end{figure}

We consider the flexural motion of the continuous beam, where its deformation is described by rotations at the points of supports and deflections between the supports. The rotational angles at the supports are denoted by $\lvert\theta\rangle\equiv (\theta_1$, $\theta_2$, \dots, $\theta_{2N})^T$. 
In what follows we shall see that the deflection of the continuous beam with infinitely many degrees of freedom can be expressed via the discrete rotational angles at the supports (joints).
The transverse deflection of each beam segment 
is taken as $u(x,t) = \Real\{\phi(x) \exp{(-\mathrm{i}\omega t)}\}$ (where $\omega$ represents frequency) in a time-harmonic form. In terms of the equation of motion of the Euler--Bernoulli beam theory,
\begin{equation}
	EI \frac{\partial^4 u}{\partial x^4} + m \frac{\partial^2 u}{\partial t^2} =0,
\end{equation}
the deflection curve $\phi(x)$ satisfies \citep{vibrationbeam}
\begin{equation}
	\label{eq8}
	\phi^{(4)}(x)-\frac{\omega^2 m}{EI}\phi(x)=0,
\end{equation}
where $m$ is the linear density, $E$ is the Young modulus, and $I$ is the moment of inertia of the cross-section. The general solution of Eq.~\eqref{eq8} is
\begin{equation}
	\label{eqa1}
	\phi(x)=C_1\sin\beta x+C_2\cos\beta x+C_3\sinh \beta x+C_4 \cosh \beta x,
\end{equation}
where
\begin{equation}
	\beta^4 \equiv \omega^2\frac{m}{EI} \quad\text{($\beta>0$)}.
\end{equation}
The deflections of the beam at the joints are zero, and the rotational angles are continuous. Thus, for a generic beam segment with a end at joint $n\in \{1,\dots,2N\}$ (which denotes the number of the joint counting from left), when the origin of the coordinate
system is set at joint $n$, the deflection function $\phi(x)$ and its
derivative satisfy the conditions
\begin{equation}
	\label{eq9}
	\phi(0)=0,\quad \phi(l)=0,\quad \phi'(0)=\theta_n,\quad \phi'(l)=\theta_j.
\end{equation}
Here, the length $l$  represents $l_{1}$ ($l_{2}$) for $j=n-1$ ($j=n+1$) when $n$ is even, and  $l_{2}$ ($l_{1}$) for $j=n-1$ ($j=n+1$) when $n$ is odd. Using Eq.~\eqref{eqa1}, we can solve $C_{1}$, $C_{2}$, $C_{3}$, $C_{4}$, but we only need $C_{4}$ in the sequel:
\begin{equation}
	\label{eq13}
	C_4(\theta_j)=\frac{1}{2\beta} \left[ \frac{B(\beta l)}{A(\beta l)}\theta_n-\frac{C(\beta l)}{A(\beta l)}\theta_j \right],
\end{equation}
where
\begin{align}
	\label{eqa2}
	A(\beta l)&\equiv 1-\cosh\beta l\cos\beta l, \\
	\label{eqa3}
	B(\beta l)&\equiv \sinh\beta l\cos\beta l-\cosh\beta l\sin\beta l, \\
	\label{eqa4}
	C(\beta l)&\equiv \sinh\beta l-\sin\beta l.
\end{align}
Thus, the deflection $\phi$ is expressed with $\lvert\theta\rangle$; and the bending moment at the $n$-th joint is given as $M_n^j=EI\phi''(0)=2EI\beta^2C_4(\theta_j)$.

Finally, the rotational angles $\lvert\theta\rangle$ at the joints are solved, by the balance equation of the bending moments at each joint,
\begin{equation}
	\sum_{j\in\{n-1, n+1\} } M_n^j=0,
\end{equation}
which is expressed as
\begin{equation}
	\label{eq17}
		\left[\frac{B(\beta l_1)}{A(\beta l_1)}+\frac{B(\beta l_2)}{A(\beta l_2)}\right]\theta_n-\frac{C(\beta l_{2(1)})}{A(\beta l_{2(1)})}\theta_{n-1}-\frac{C(\beta l_{1(2)})}{A(\beta l_{1(2)})}\theta_{n+1}=0.
\end{equation}
When $n$ is odd, the coefficients $C$ and $A$ before $\theta_{n-1(n+1)}$ in Eq.~\eqref{eq17} are both functions of $\beta l_{2(1)}$; when $n$ is even, they are functions of $\beta l_{1(2)}$. Under fixed-end boundary conditions shown in Fig.~1(a), $\theta_{0,2N+1}=0$ should be substituted into Eq.~\eqref{eq17} for the first and ($2N$)-th joints. Now, Eq.~\eqref{eq17}, representing a set of $2N$ equations, can be rewritten into a matrix form
\begin{equation}
	\label{eqa5}
	H^\mathrm{beam}\vert \theta \rangle =0,
\end{equation}
where $H^\mathrm{beam}$ is the dynamic matrix. The main-diagonal elements of $H^\mathrm{beam}$ are $\left[\frac{B(\beta l_1)}{A(\beta l_1)}+\frac{B(\beta l_2)}{A(\beta l_2)}\right]$, and the subdiagonal elements of $H^\mathrm{beam}$ are staggered numbers of $-\frac{C(\beta l_{1})}{A(\beta l_{1})}$ and $-\frac{C(\beta l_{2})}{A(\beta l_{2})}$. The existence of a nontrivial solution $\vert \theta \rangle$ is equivalent to that of \emph{a zero eigenvalue} for matrix $H^\mathrm{beam}$, and the frequencies $\beta$ of zero eigenvalues in this regard correspond to the natural frequencies of a finite continuous beam structure.

The dynamic matrix $H^\mathrm{beam}$ is analogous to the Hamiltonian matrix of the Su--Schrieffer--Heeger (SSH) chain \citep{SSH1979,Asbth2016} whose topological dynamics has been known for a long time, and the rotational angles of joints are analogous to the wavefunction at the sites. Thus, with nontrivial topological properties, the natural frequencies of a finite structure will extend beyond the frequency range of the bulk bands of a boundary-free structure, manifested as the emergence of topological states. However, such a similarity in the matrix form is present only when the most important parameter, the frequency $\beta$, is kept fixed; on the contrary, since the dynamic matrix $H^\mathrm{beam}$ constantly varies with $\beta$, the topological properties of continuous beam structures are very different from the discrete system. As a result, our continuous beam structures have \emph{vibration modes in multiple frequency ranges} as shown in Fig.~1(b)--(d) (the natural frequencies $\beta$ are piecewise distributed in intervals across the full spectrum, in striking contrast to the discrete system), due to the \emph{infinitely many degrees of freedom} in the continuum. We prove the existence of topological states in the following.

\section{Existence of topological edge states within bandgaps}
\label{sec2}
\subsection{Existence of topological edge states}
Since the dynamic matrix of the continuous beam has chiral symmetry \citep{Schnyder2008,Chiu2016,Asbth2016,Maffei2018} (up to subtraction of a certain multiple of the identity matrix), the frequencies $\beta_t$ of the topological edge states must correspond to the case that the main-diagonal elements of the dynamic matrix $H^\mathrm{beam}$ are zero, that is,
\begin{equation}
	\label{eq18}
	\frac{B(\beta_t l_1)}{A(\beta_t l_1)}+\frac{B(\beta_t l_2)}{A(\beta_t l_2)}=0.
\end{equation}
From Eq.~\eqref{eq18}, we solve the frequencies $\beta_t$ where the topological states may occur. To this end, we need to prove the existence of positive
roots of Eq.~\eqref{eq18}.

To prove the existence of topological states,  we examine the properties of the function $\frac{B(\beta l)}{A(\beta l)}$ when $l$ is fixed. Noting that the function $\frac{B(\beta l)}{A(\beta l)}$ is continuously differentiable with respect to $\beta$ when the denominator $A(\beta l)\neq 0$, we have
\begin{equation}
	\label{eq19}
	\frac{1}{l}\frac{d}{d\beta}\left[\frac{B(\beta l)}{A(\beta l)}\right]=\frac{B'\cdot A-B\cdot A'}{A^2}=\frac{(\sinh \beta l-\sin \beta l)^2}{A^2}=\frac{C^2}{A^2}> 0,
\end{equation}
where $A'=-B$ (the prime denotes differentiation with respect to the argument $\beta l$), indicating that the function $\frac{B(\beta l)}{A(\beta l)}$ increases monotonically with $\beta$ in any interval in which $A(\beta l)\neq 0$. Since the roots $\beta_0$ of $A(\beta l)=1-\cosh\beta l\cos\beta l=0$ satisfy $\beta_0 l>3\pi/2$, we have $\frac{1}{\cos \beta_0 l} = \cosh \beta_0 l > \sqrt{2}$, i.e., $\lvert\tan \beta_0 l \rvert>1$. And since $-1<\tanh \beta_0 l<1$, we have
\[ B(\beta_0 l)=\cos\beta_0 l\sinh\beta_0 l-\sin\beta_0 l\cosh\beta_0 l=\tanh \beta_0 l-\tan \beta_0 l\neq 0, \]
that is, the zeros of the function $A(\beta l)$ do not coincide with the zeros of $B(\beta l)$. Therefore, when $A(\beta_0 l)=0$, the values of $B(\beta_0 l)$ are finite, and hence $\frac{B(\beta_0 l)}{A(\beta_0 l)}\rightarrow \infty$.

Next, we examine the properties of $\left[\frac{B(\beta l_1)}{A(\beta l_1)}+\frac{B(\beta l_2)}{A(\beta l_2)}\right]$ as a function of $\beta$. The $\beta_0$ which render the denominators $A(\beta_0 l_1)=0$ and $A(\beta_0 l_2)=0$ divide the frequencies $\beta$ into multiple intervals; and at each root $\beta_0$, $\frac{B(\beta_0 l_1)}{A(\beta_0 l_1)}+\frac{B(\beta_0 l_2)}{A(\beta_0 l_2)} \to \infty$. From Eq.~\eqref{eq19}, we know that the function $\left[\frac{B(\beta l_1)}{A(\beta l_1)}+\frac{B(\beta l_2)}{A(\beta l_2)}\right]$ increases monotonically in each interval $(\beta_0^{(n-1)},\beta_0^{(n)})$, that is, in each interval, the function $\left[\frac{B(\beta l_1)}{A(\beta l_1)}+\frac{B(\beta l_2)}{A(\beta l_2)}\right]$ increases monotonically from $-\infty$ to $+\infty$; thus it must have one and only one root $\beta_t^{(n)}$ in the interval $(\beta_0^{(n-1)},\beta_0^{(n)})$.

Finally, we give an example to graphically illustrate the above results. The structure under the cyclic boundary condition (i.e., Born--von Karman boundary condition) is considered, where we take $\theta_{0}=\theta_{2N}$ and $\theta_{2N+1}=\theta_{1}$ in the balance equations in Eq.~\eqref{eq17} of the first and $(2N)$-th joints. Similarly, Eq.~\eqref{eq17} can be rewritten into a matrix equation $H^\mathrm{beam}_\text{B--K}\vert \theta \rangle =0$, whose solutions $\beta$ correspond to the frequency bands of a boundary-free continuous beam structure. Under the Born--von Karman boundary condition, with parameters $l_1=40~\mathrm{mm}$, $l_2=50~\mathrm{mm}$, and $N=16$, Fig.~2(a) depicts all the eigenvalues $y$ (ordinate) of the matrix $H^\mathrm{beam}_\text{B--K}$ as $\beta$ (abscissa) varies continuously. The natural frequencies $\beta$ of the periodic continuous beam structure (i.e., the solutions of $H^\mathrm{beam}_\text{B--K}\vert \theta \rangle =0$) are displayed in Fig.~2(a) as all intersection points of the black curves and the gray straight line $y=0$, which are piecewise distributed across the spectrum. All vertical lines correspond to $\beta=\beta_0^{(n)}$, roots of either $A(\beta_0 l_1)=0$ or $A(\beta_0 l_2)=0$. There is only one blue dashed curve between every two vertical lines, which represents the value of the function $\left[\frac{B(\beta l_1)}{A(\beta l_1)}+\frac{B(\beta l_2)}{A(\beta l_2)}\right]$. It can be seen that the blue curve increases monotonically from $-\infty$ to $+\infty$ within each interval $(\beta_0^{(n-1)},\beta_0^{(n)})$, and the intersections of the blue curve with $y=0$ correspond to the solution of Eq.~\eqref{eq18} (i.e., $\beta_t^{(n)}$). Thus we can see from Fig.~2(a) that there exists one and only one frequency $\beta_t^{(n)}$ (which belongs to the set $\{\beta_t\}$ containing the frequencies of the topological edge states) within each interval $(\beta_0^{(n-1)},\beta_0^{(n)})$ for the continuous beam structure.

The above results can be summarized into a theorem:
\begin{theorem}
	\label{theorem1}
	The frequencies of the topological edge states of a simply supported continuous beam with alternating spans $l_1$ and $l_2$ belong to the set $\{\beta_t\}$ of all the positive roots of $\frac{B(\beta l_1)}{A(\beta l_1)} + \frac{B(\beta l_2)}{A(\beta l_2)} = 0$, where $A(\beta l)=1-\cosh\beta l\cos\beta l, \ \
	B(\beta l)=\sinh\beta l\cos\beta l-\cosh\beta l\sin\beta l$.
	There is one $\beta_t$ in each interval $(\beta_0^{(n-1)}, \beta_0^{(n)})$ of the roots of $A(\beta l_1) A(\beta l_2) = 0$.
\end{theorem}

For the set $\{\beta_t\}$ containing all the frequencies of the topological edge states, we divide it into two subsets $\{\beta_{t{<}}\}$ and $\{\beta_{t>}\}$, where $\{\beta_t\}=\{\beta_{t<}\}\cup\{\beta_{t>}\}$. $\{\beta_{t<}\}$ is defined as the subset containing $\beta_{t<}^{(n)}$ for which the subdiagonal elements of the dynamic matrix satisfies
\begin{equation}
	\label{eq20}
	\bigg \vert\frac{C(\beta_{t} l_1)}{A(\beta_{t} l_1)}\bigg \vert<\bigg \vert\frac{C(\beta_{t} l_2)}{A(\beta_{t} l_2)}\bigg \vert,
\end{equation}
and $\{\beta_{t>}\}$ is defined as the subset which contains $\beta_{t>}^{(n)}$ satisfying
\begin{equation}
	\label{eq21}
	\bigg \vert\frac{C(\beta_{t} l_1)}{A(\beta_{t} l_1)}\bigg \vert>\bigg \vert\frac{C(\beta_{t} l_2)}{A(\beta_{t} l_2)}\bigg \vert.
\end{equation}
The topological states exist at every $\beta_{t<}^{(n)}$ in set $\{\beta_{t<}\}$, where the rotational angles at the joints and the deflections of the beam segments reach the maxima around the edges; there are no topological states at frequencies in set $\{\beta_{t>}\}$. The case when $\beta_t$ satisfies $\Big \lvert\frac{C(\beta_t l_1)}{A(\beta_t l_1)}\Big \rvert=\Big \lvert\frac{C(\beta_t l_2)}{A(\beta_t l_2)}\Big \rvert$ corresponds to the topological phase transition point.

We illustrate that topological states exist and only exist at frequency $\beta_t^{(n)}$ when $\Big \vert\frac{C(\beta_t l_1)}{A(\beta_t l_1)}\Big \vert<\Big \vert\frac{C(\beta_t l_2)}{A(\beta_t l_2)}\Big \vert$, with an exampled continuous beam structure with parameters $l_1=40~\mathrm{mm}$, $l_2=50~\mathrm{mm}$, and $N=16$. As shown in Fig.~2(b), $\Big \vert\frac{C(\beta_t l_1)}{A(\beta_t l_1)}\Big \vert<\Big \vert\frac{C(\beta_t l_2)}{A(\beta_t l_2)}\Big \vert$ is satisfied at $\beta_t^{(1)}$, $\beta_t^{(3)}$, and $\beta_t^{(5)}$; thus we find degenerate topological states at these frequencies as shown in Fig.~1(b), within certain intervals such as $(\beta_0^{(0)},\beta_0^{(1)})$, $(\beta_0^{(2)},\beta_0^{(3)})$, $(\beta_0^{(4)},\beta_0^{(5)})$.  When the parameters are exchanged and taken as $l_1=50~\mathrm{mm}$, $l_2=40~\mathrm{mm}$, as shown in Fig.~1(c), topological edge states occur in the intervals where topological states are not seen in Fig.~1(b).
\begin{figure}[tb!]
	\centering
	\includegraphics[width=\linewidth]{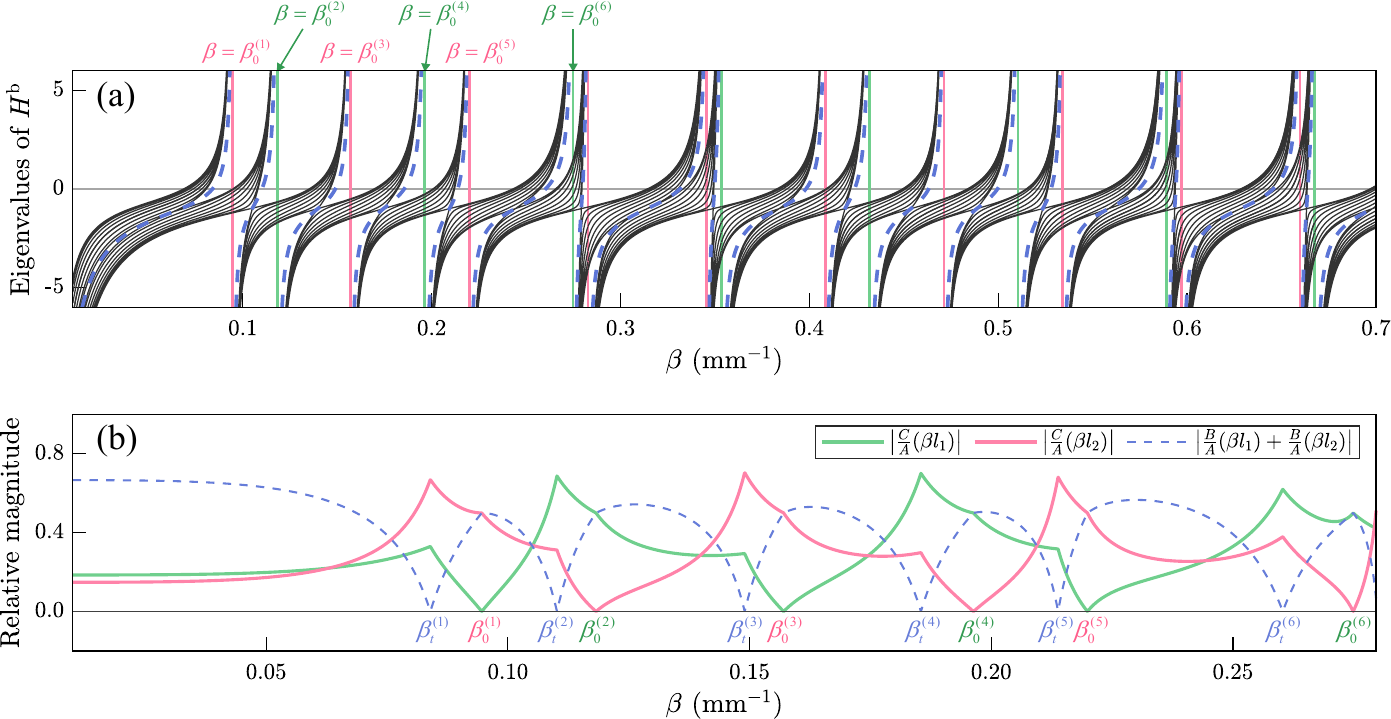}
	\caption{ (a) Bulk bands of the boundary-free continuous beam structure (corresponding to all intersections of the black curves with the horizontal gray straight line) and frequencies at which topological states may exist (the intersections of the blue dashed curves with the horizontal gray straight line). The pink (green) vertical lines represent the roots of the equation $A(\beta_0 l_{2(1)})=0$. (b) Relative magnitudes of subdiagonal elements $\frac{C}{A}(\beta l_1)$ and $\frac{C}{A}(\beta l_2)$ for the continuous beam structure with $l_1=40~\mathrm{mm}$ and $l_2=50~\mathrm{mm}$. The frequency $\beta_t^{(n)}$ in each interval $(\beta_0^{(n-1)},\beta_0^{(n)})$ belongs to the set $\{\beta_t\}$ containing the frequencies of topological states; an edge state exists if and only if $\big \vert\frac{C}{A}(\beta_t l_1)\big \vert<\big \vert\frac{C}{A}(\beta_t l_2)\big \vert$ at $\beta_t^{(n)}$. } \label{fig:2}
\end{figure}

\subsection{Existence of frequency bands and topological states within bandgaps}
Under Bloch periodic boundary conditions, the dispersion relation of the continuous beam is given in Fig.~1(d). By comparing the frequencies of the bulk bands and topological states in Fig.~1(b)--(d), when no band crossings occur (i.e., $\Big \lvert\frac{C(\beta_t l_1)}{A(\beta_t l_1)}\Big \rvert\neq\Big \lvert\frac{C(\beta_t l_2)}{A(\beta_t l_2)}\Big \rvert$), we have the following theorem (the proof is given below):
\begin{theorem}
	\label{theorem2}
	There is one frequency band between two consecutive frequencies $(\beta_t^{(n-1)},\beta_t^{(n)})$ of the set $\{\beta_t\}$.
\end{theorem} The bulk band between $(\beta_t^{ (n-1)},\beta_t^{(n)})$ is referred to as the $n$-th frequency band, where $\beta_t^{(0)}\equiv 0$; the bandgap between the $n$-th and $(n+1)$-th frequency bands is referred to as the $n$-th bandgap. Theorem~\ref{theorem2} implies that all the topological states exist within the bandgaps.

Then, we prove Theorem~\ref{theorem2}. We first determine the locations of the bulk bands by an infinite Bloch-periodic continuous beam structure. For an infinite continuous beam structure, the periodic unit cell is shown in Fig.~1(a), and the rotational angles of the joints at sublattices A and B in the $m$-th unit cell are $\theta_\mathrm{A} ^m$ and $\theta_\mathrm{B}^m$. The dynamic equations are
\begin{align}
	\label{eq22}
	\left[\frac{B(\beta l_1)}{A(\beta l_1)}+\frac{B(\beta l_2)}{A(\beta l_2)}\right]\theta_\mathrm{A}^m-\frac{C(\beta l_2)}{A(\beta l_2)}\theta_\mathrm{B}^{m-1}-\frac{C(\beta l_1)}{A(\beta l_1)}\theta_\mathrm{B}^{m}&=0,\\
	\label{eq23}
	\left[\frac{B(\beta l_1)}{A(\beta l_1)}+\frac{B(\beta l_2)}{A(\beta l_2)}\right]\theta_\mathrm{B}^m-\frac{C(\beta l_1)}{A(\beta l_1)}\theta_\mathrm{A}^{m}-\frac{C(\beta l_2)}{A(\beta l_2)}\theta_\mathrm{A}^{m+1}&=0.
\end{align}
Substituting Bloch's theorem $\theta_{\mathrm{A(B)}}^{m+1} = \theta_{\mathrm{A(B)}}^{m} \exp{(\mathrm{i}kL)}$ into Eqs.~\eqref{eq22} and \eqref{eq23}, we obtain
\begin{equation}
	\label{eq24}
	H_{\mathrm{Bloch}}^\mathrm{beam}\vert \theta\rangle=
	\begin{bmatrix}
		\displaystyle \frac{B(\beta l_1)}{A(\beta l_1)}+\frac{B(\beta l_2)}{A(\beta l_2)} & \displaystyle -\frac{C(\beta l_1)}{A(\beta l_1)}-\frac{C(\beta l_2)}{A(\beta l_2)} \exp{(-\mathrm{i}kL)}\\[9pt]
		\displaystyle -\frac{C(\beta l_1)}{A(\beta l_1)}-\frac{C(\beta l_2)}{A(\beta l_2)} \exp{(\mathrm{i}kL)} & \displaystyle
		\frac{B(\beta l_1)}{A(\beta l_1)}+\frac{B(\beta l_2) }{A(\beta l_2) }
	\end{bmatrix}
	\begin{bmatrix}
		\theta_\mathrm{A}^m\\ \theta_\mathrm{B}^m
	\end{bmatrix}=0,
\end{equation}
where $L=l_1+l_2$ is the lattice constant. For Eq.~\eqref{eq24} to have nontrivial solutions, it must hold that
\begin{equation}
	\label{eq25}
	\frac{B(\beta l_1)}{A(\beta l_1)}+\frac{B(\beta l_2)}{A(\beta l_2)}=\pm\sqrt{\left[\frac{C(\beta l_1)}{A(\beta l_1)}\right]^2+\left[\frac{C(\beta l_2)}{A(\beta l_2)}\right]^2+2\frac{C(\beta l_1)}{A(\beta l_1)}\frac{C(\beta l_2)}{A(\beta l_2)}\cos kL}.
\end{equation}
Solutions of Eq.~\eqref{eq25} correspond to the range of the frequency bands, which are invariant under an exchange of $l_1$ and $l_2$. The dispersion relation solved from Eq.~\eqref{eq25} is shown in Fig.~1(d). To prove that the topological states do not overlap with the range of bulk bands, Eq.~\eqref{eq25} is rewritten into
\begin{align}
	\label{eq26}
	\left[\frac{B(\beta l_1)}{A(\beta l_1)}+\frac{B(\beta l_2)}{A(\beta l_2)}\right]^2&=\left[\frac{C(\beta l_1)}{A(\beta l_1)}+\frac{C(\beta l_2)}{A(\beta l_2)}\right]^2+\frac{2C(\beta l_1)C(\beta l_2)}{A(\beta l_1)A(\beta l_2)}(\cos kL-1),\\
	\label{eq27}
	\left[\frac{B(\beta l_1)}{A(\beta l_1)}+\frac{B(\beta l_2)}{A(\beta l_2)}\right]^2&=\left[\frac{C(\beta l_1)}{A(\beta l_1)}-\frac{C(\beta l_2)}{A(\beta l_2)}\right]^2+\frac{2C(\beta l_1)C(\beta l_2)}{A(\beta l_1)A(\beta l_2)}(\cos kL+1).
\end{align}
Since $\frac{B}{A}(\beta_t l_1)+\frac{B}{A}(\beta_t l_2)=0$ holds at the frequencies of the topological states, we substitute it into Eqs.~\eqref{eq26} and \eqref{eq27}, rendering the left-hand sides both equal to 0. When $ \big\lvert\frac{C}{A}(\beta_t l_1) \big\rvert\neq \big\lvert\frac{C}{A}(\beta_t l_2) \big\rvert$, if $\frac{C(\beta_t l_1)C(\beta_t l_2)}{A(\beta_t l_1)A(\beta_t l_2)}\leq 0$, the right-hand side of Eq.~\eqref{eq26} is larger than 0; if $\frac{C(\beta_t l_1)C(\beta_t l_2)}{A(\beta_t l_1)A(\beta_t l_2)}>0$, the right-hand side of Eq.~\eqref{eq27} becomes larger than 0. In conclusion, under the condition of $ \big\lvert\frac{C}{A}(\beta_t l_1) \big\rvert\neq \big\lvert\frac{C}{A}(\beta_t l_2) \big\rvert$, the frequencies corresponding to the topological states are not within the solution range of the periodic frequency bands (that is, the frequencies $\beta=\beta_t$ are not the solutions of Eq.~\eqref{eq25} for any value of $kL$), so the topological states do not overlap with the frequency bands. We note that the exceptional points of $ \big\lvert\frac{C}{A}(\beta_t l_1) \big\rvert = \big\lvert\frac{C}{A}(\beta_t l_2) \big\rvert$ in rare cases correspond to the Dirac points where two bulk bands touch, closing the bandgap.

Next, we continue to prove that there must be a frequency band between every two consecutive frequencies $(\beta_t^{(n-1)},\beta_t^{(n)})$ of the set $\{\beta_t\}$ containing the frequencies of the topological states, that is, the solutions of Eq.~\eqref{eq25} exist in each interval $(\beta_t^{(n-1)},\beta_t^{(n)})$ for any $k$ in the Brillouin zone. To this end, we define a function
\begin{equation}
	\label{eq28}
	f(\beta)\equiv \left[\frac{C(\beta l_1)}{A(\beta l_1)}\right]^2+\left[\frac{C(\beta l_2)}{A(\beta l_2)}\right]^2-\left[\frac{B(\beta l_1)}{A(\beta l_1)}\right]^2-\left[\frac{B(\beta l_2)}{A(\beta l_2)}\right]^2 +\frac{2C(\beta l_1)C(\beta l_2)}{A(\beta l_1)A(\beta l_2)}\cos kL-\frac{2B(\beta l_1)B(\beta l_2)}{A(\beta l_1)A(\beta l_2)},
\end{equation}
which is simply the difference of the two sides of Eq.~\eqref{eq26} (or \eqref{eq27}). From the analysis in the previous paragraph, we know that $f(\beta_t)>0$ and takes a finite value when $ \big\lvert\frac{C}{A}(\beta_t l_1) \big\rvert\neq \big\lvert\frac{C}{A}(\beta_t l_2) \big\rvert$. Meanwhile, from Theorem~\ref{theorem1}, $\beta_0^{(n)}$ which satisfies either $A(\beta_0^{(n)}l_1)=0$ or $A(\beta_0^{(n)}l_2)=0$ must exist in the interval $(\beta_t^{(n-1)},\beta_t^{(n)})$. Since
\begin{equation}
	\label{eq29}
	C^2(\beta l)-B^2(\beta l)=-2\sinh(\beta l)\sin(\beta l)\cdot A(\beta l),
\end{equation}
considering the behavior of $f(\beta)$ around $\beta=\beta_0^{(n)}$ when $A(\beta l_1)\rightarrow 0^{\pm}$ (the case $A(\beta l_2)\rightarrow 0^{\pm}$ is analogous), we have
\begin{equation}	
	\label{eq30}
	\begin{split}
		f(\beta) &=  \left[\frac{C(\beta l_1)}{A(\beta l_1)}\right]^2-\left[\frac{B(\beta l_1)}{A(\beta l_1)}\right]^2 + \left[ \frac{2C(\beta l_1)C(\beta l_2)\cos kL-2B(\beta l_1)B(\beta l_2)}{A(\beta l_2)}\right] \frac{1}{A(\beta l_1)} + O(1)\\
		&=\frac{1}{A(\beta l_1)}\left[-2\sinh(\beta l_1)\sin(\beta l_1)+\frac{2C(\beta l_1)C(\beta l_2)\cos kL-2B(\beta l_1)B(\beta l_2)}{A(\beta l_2)}\right] + O(1) ,
	\end{split}
\end{equation}
i.e., $f(\beta) \sim (\beta-\beta_0^{(n)})^{-1}$, and thus $f(\beta)\rightarrow -\infty$ in either of the one-sided limits $\beta\rightarrow\beta_0^{(n)\pm}$. In terms of this result and the condition $f(\beta_t)>0$, for each given $kL\in(-\pi,\pi]$, there must exist a zero point of $f(\beta)=0$ between the frequencies $(\beta_t^{(n-1)},\beta_t^{(n)})$, that is, Eq.~\eqref{eq25} must have a solution between the interval $(\beta_t^{(n-1)},\beta_t^{(n)})$, and hence there must be a frequency band between the adjacent frequencies of the topological states. Through the above proofs, we find that the topological states, whose frequencies may only take the values in $\{\beta_t\}$, all exist within the bandgaps and do not overlap with the frequency bands. Theorem~2 is proved. The spectra obtained from the example structures -- Fig.~1(b),(c) -- also show the existence of frequency bands between the topological states and the phenomenon that the topological states and frequency bands do not overlap.

\subsection{Localization of topological edge states}
\begin{figure}[tb!]
	\centering
	\includegraphics[width=0.8\linewidth]{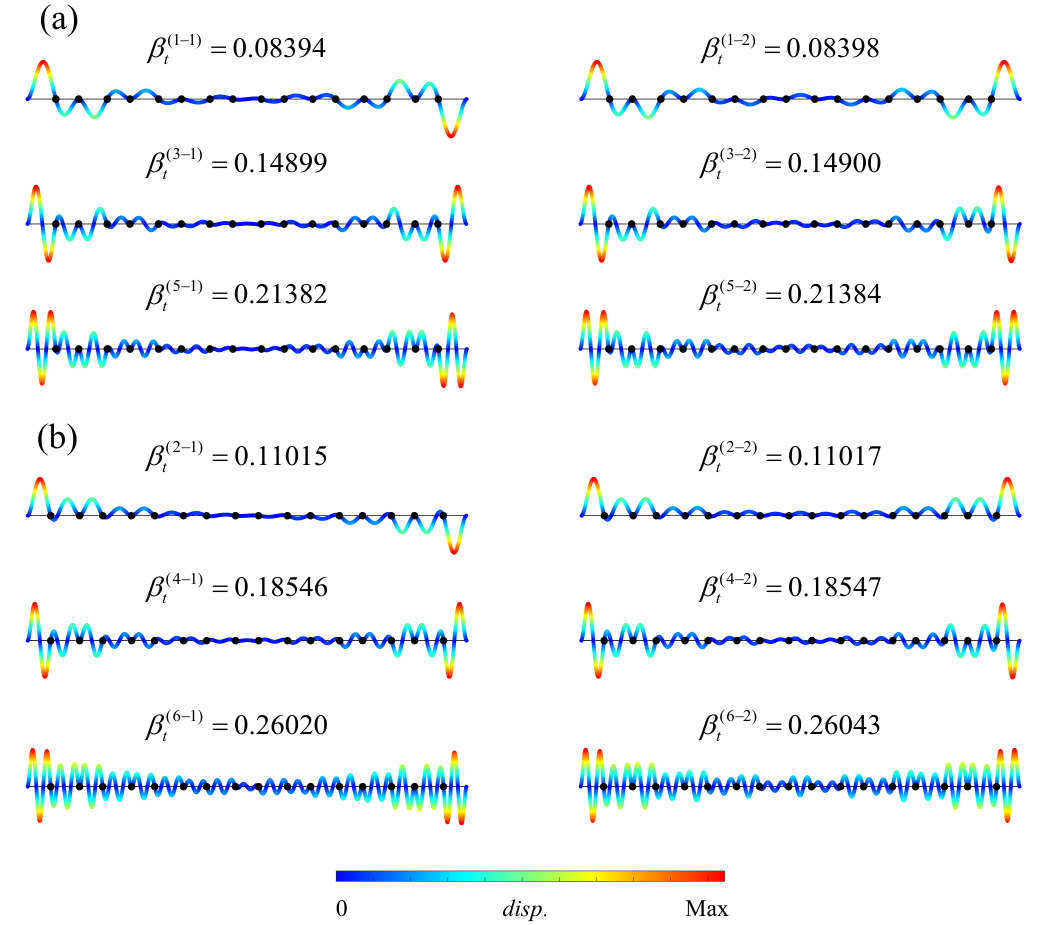}
	\caption{Topological localized edge modes at the first-, second- and third-order modes of the continuous beam with $l_1=40~\mathrm{mm}$ and $l_2=50~\mathrm{mm}$ (a), $l_1=50~\mathrm{mm}$ and $l_2=40~\mathrm{mm}$ (b).} \label{fig:3}
\end{figure}
We demonstrate the vibration modes of the localized topological edge states indicated in Figs.~1(b) and (c). For the continuous beam structure with parameters $l_1=40~\mathrm{mm}$, $l_2=50~\mathrm{mm}$, and $N=8$, the topological edge modes ($\beta=\beta_t^{(1)}$) in the first bandgap are shown on the top of Fig.~3(a), where each segment of the beam between two adjacent simple supports vibrates predominantly at its first-order mode shape. Due to the finite number of unit cells, there are two almost degenerate topological states at $\beta=\beta_t^{(1)}$ (represented by $\beta=\beta_t^{(1\text{--}1)}$ and $\beta_t^{(1\text{--}2)}$), and the rotational angles are symmetrically and antisymmetrically distributed, which are localized at the leftmost and rightmost joints. The deflections are also localized at the leftmost and rightmost beam segments, shown in Fig.~3(a). The topological edge modes in the third and fifth bandgaps ($\beta=\beta_t^{(3)}$ and $\beta=\beta_t^{(5)}$) are also shown in Fig.~3(a), where each segment is at its second- and third-order mode shapes. Similarly, for the continuous beam structures with parameters $l_1=50~\mathrm{mm}$, $l_2=40~\mathrm{mm}$, and $N=8$, the topological edge states in the second, fourth, and sixth bandgaps ($\beta=\beta_t^{(2)}$, $\beta_t^{(4)}$, $\beta_t^{(6)}$) are shown in Fig.~3(b), where the segments  vibrate at their first-, second-, and third-order mode shapes. These theoretical results are verified by numerical simulations in Subsection~3.4.

We illustrate the localization characteristics of the topological state at the left boundary, while the right boundary is similar. Now that $\beta = \beta_t$, we apply Eq.~\eqref{eq17} for $n = 1$, which yields
\begin{equation}
	\theta_2 = 0
\end{equation}
(note that we have used Eq.~\eqref{eq18}); meanwhile, substituting $n = 3$, $5$, \dots\ into Eq.~\eqref{eq17} gives the relations for the even-numbered joints:
\begin{equation}
	-\frac{C(\beta_t l_{2})}{A(\beta_t l_{2})}\theta_{2n}-\frac{C(\beta_t l_{1})}{A(\beta_t l_{1})}\theta_{2n+2}=0 .
\end{equation}
This implies that the rotational angles are zero at the even-numbered joints for the topological state at the left boundary. For the odd-numbered joints, again from Eq.~\eqref{eq17}, we know that the rotational angles reach the maximum value at the first joint and decay exponentially away from the left edge with $\theta_{ n+2}=c_\mathrm{decay}\cdot\theta_{n}$, where
\begin{equation}
	c_\mathrm{decay}=-\frac{C(\beta_t l_1) A(\beta_t l_2)}{C(\beta_t l_2 )A(\beta_t l_1)}
\end{equation}
and $\lvert c_\mathrm{decay}\rvert<1$.

The localization is also manifested in the deflections of the beam segments as shown in Fig.~3(a),(b). We show the localization at the left boundary in the beam segments with length $l_1$ as an example. The joint rotational angles $\theta_3$ and $\theta_4$ are just a rescaling of the angles $\theta_1$ and $\theta_2$ (note that $\theta_{2n}$ is essentially zero near the left boundary). Because Eq.~\eqref{eq9} is linear with respect to $C_{1,2,3,4}$ and $\theta_{n,j}$, the deflection of the fourth beam segment is also a simple rescaling of the deflection of the second segment, with the same scaling coefficient $c_\mathrm{decay}$ as that of the angles, $\theta_3/\theta_1$. For all segments with length $l_1$, the beam deflections reach the maximum value at the beam between the first and second joints and decay exponentially away from the left edge. The case for the beam segments with length $l_2$ is similar.

\subsection{Numerical results of topological edge states of continuous beam structures}
We present detailed numerical results of the topological edge states, verifying the theoretical conclusions. We construct a continuous beam structure using the beam module in the structural mechanics branch of COMSOL Multiphysics. The material is structural steel, with elastic modulus $200\times 10^9~\mathrm{Pa}$, density $7850~\mathrm{kg/m^3}$, and Poisson's ratio 0.3. The cross-sections are all selected as rectangular sections with width and height of 1~mm. The total length of the continuous beam is consistent with the theoretical model where $N=8$.
\begin{figure}[tpb!]
	\centering
	\includegraphics[width=0.7\linewidth]{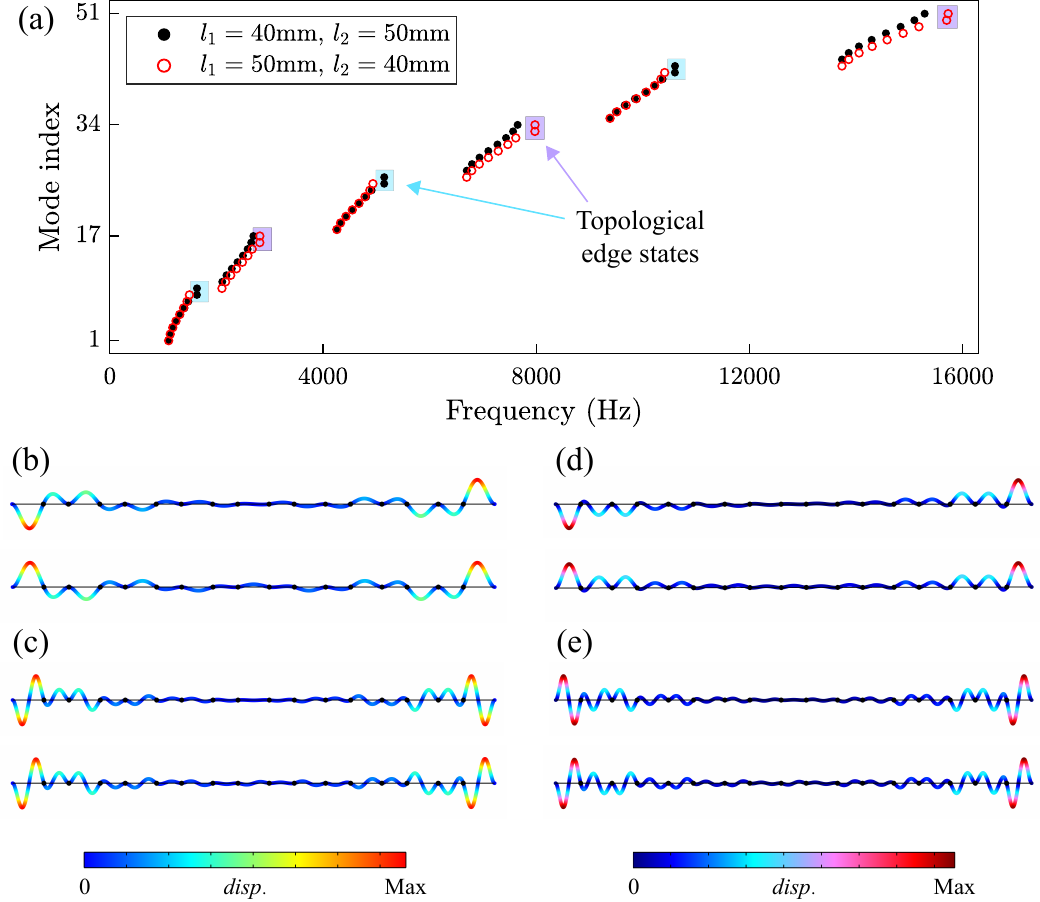}
	\caption{Numerical results of topological states in continuous beam structure. (a) Comparison of spectra of continuous beam structures with different (interchanged) parameters $l_1$ and $l_2$. The black solid (red hollow) dots represent the spectrum of the continuous beam structure with $l_1=40\ (50)~\mathrm{mm}$ and $l_2=50\ (40)~\mathrm{mm}$. (b),(c) Topological states with the first- and second-order mode shapes of the structure with $l_1=40~\mathrm{mm}$ and $l_2=50~\mathrm{mm}$. The maximum absolute values of rotational angles are obtained at the leftmost and rightmost joints. (d),(e) Topological states of the structure with $l_1=50~\mathrm{mm}$ and $l_2=40~\mathrm{mm}$.} \label{fig:app}
\end{figure}

First, the length intervals of the simple supports in the continuous beam are set to $l_1=40~\mathrm{mm}$ and $l_2=50~\mathrm{mm}$, and then the natural frequencies of the continuous beam structure are calculated. As shown in Fig.~4(a), the black solid dots represent the spectrum composed of 51 modes at the lowest frequencies of the structure. The first-order vibration modes of the continuous beam structure (whose beam segment between every two supports are at their first-order mode shapes) correspond to the 1st--17th modes, of which the 8th mode with a frequency of 1544.6~Hz and the 9th mode with a frequency of 1560.1~Hz correspond to the topological edge states, and the rotational angles of the topological states with first-order vibration modes are localized at the leftmost and rightmost joints of the continuous beam as shown in Fig.~4(b). The second-order vibration modes correspond to the 18th--34th modes, of which the 25th mode with a frequency of 4966.2~Hz and the 26th mode with a frequency of 4971.4~Hz correspond to the topological edge states, and the rotational angles of the topological state with second-order vibration modes are also localized as shown in Fig.~4(c). The pattern of the third-order vibration modes is similar, and the topological states appear at frequencies 10270~Hz and 10272~Hz. As shown in the blue boxes in Fig.~4(a), the two topological states with the first-order vibration modes are within the first bandgap, and the two topological states with the second-order (third-order) vibration modes appear in the third (fifth) bandgap.

Then, the length intervals of the simple supports in the continuous beam are interchanged and set to $l_1=50~\mathrm{mm}$ and $l_2=40~\mathrm{mm}$, and the natural frequencies are calculated. As shown in Fig.~4(a), the red hollow dots represent the spectrum composed of 51 modes at the lowest frequencies. The first-order vibration modes  correspond to the 1st--17th modes, of which the 16th mode with a frequency of 2353.8~Hz and the 17th mode with a frequency of 2358.2~Hz correspond to the localized topological edge states, shown in Fig.~4(d). The second-order vibration modes correspond to the 18th--34th modes, of which the 33th mode with a frequency of 6578.5~Hz and the 34th mode with a frequency of 6580.7~Hz correspond to the localized topological edge states, shown in Fig.~4(e). The pattern of the third-order vibration modes is similar, and the topological states appear at frequencies 13016~Hz and 13018~Hz. As shown in the purple boxes in Fig.~4(a), the two topological states with first-order vibration modes now appear in the second bandgap, and the two topological states with second-order (third-order) vibration modes appear in the fourth (sixth) bandgap.

\section{Topological index related to the Zak phase that determines the existence of edge states}
We introduce a \emph{topological index} $\mu^{(n)}$ related to the bulk Zak phase  \citep{Zak1989,Xiao2015exp,Miert2016} to determine the topological edge states, which is defined as
\begin{equation}
	\label{eqindex}
	\mu^{(n)}=\gamma^{(n)}+\pi\cdot l^{(n)} \mod 2\pi,
\end{equation}
where the Zak phase $\gamma^{(n)}$ for the $n$-th bulk band is
\begin{equation}
	\label{eqzak}
	\gamma^{(n)}=\mathrm{i}\int_{-\pi/L}^{\pi/L}dk \langle \theta^{(n)}\rvert \partial_k \lvert \theta^{(n)}\rangle \mod 2\pi,
\end{equation}
and $l^{(n)}$ denotes the number of ``\emph{eigenvalue crossing}'' occurring at positive eigenvalues within the frequency range $(\beta_t^{(n-1)},\beta_t^{(n)})$ of the $n$-th bulk band. When $\mu^{(n)}=\pi$, nontrivial topological edge states must exist in the $n$-th bandgap; when $\mu^{(n)}=0$, there are no topological edge states. The criterion Eq.~\eqref{eqindex} \emph{strictly establishes the bulk--edge correspondence} \citep{SurfaceImpedance2014,Asbth2016} and gives the same results as predicted via Eqs.~\eqref{eq20} and \eqref{eq21}. We present details and derivations in the following.

We show the rigorous relation between the Zak phase and the topological index which strictly determines the edge states.
The solution of rotational angles for eigenequation \eqref{eq24} is
\begin{equation}
	\label{eq6}
	\vert \theta \rangle =
	\begin{bmatrix}
		\theta_\mathrm{A}^m\\ \theta_\mathrm{B}^m
	\end{bmatrix}
	= \frac{1}{c_\mathrm{norm}}
	\begin{bmatrix}
		\displaystyle \frac{B(\beta l_1)}{A(\beta l_1)}+\frac{B(\beta l_2)}{A(\beta l_2)} \\[9pt]
		\displaystyle \frac{C(\beta l_1)}{A(\beta l_1)}+\frac{C(\beta l_2)}{A(\beta l_2)} \exp{(\mathrm{i}kL)}
	\end{bmatrix}
	= \frac{\sqrt{2}}{2}\begin{bmatrix} 1\\ \exp{\bigl(\mathrm{i}\chi(k)\bigr)} \end{bmatrix},
\end{equation}
where
\begin{equation}
	\chi(k)=\arg \left[ \dfrac{C(\beta l_1)A(\beta l_2)+C(\beta l_2)A(\beta l_1) \exp{(\mathrm{i}kL)}}{B(\beta l_1)A(\beta l_2)+B(\beta l_2)A(\beta l_1)} \right] .
\end{equation}
The Zak phase for each band in Fig.~1(d) is calculated from vector $\lvert \theta \rangle$ as
\begin{equation}
	\label{eq7}
	\gamma=\mathrm{i}\int_{-\pi/L}^{\pi/L}dk \langle \theta\rvert \partial_k \lvert \theta\rangle=\mathrm{i}\int_{-\pi/L}^{\pi/L}dk\left(\theta_\mathrm{A}^{m\ast} \frac{\partial{\theta_\mathrm{A}^m}}{\partial{k}}+\theta_\mathrm{B}^{m\ast}\frac{\partial{\theta_\mathrm{B}^m}}{\partial{k}}\right)=-\frac{1}{2}\chi(k)\big\vert_{-\pi/L}^{\pi/L}=-\chi(k)\big\vert_{0}^{\pi/L}.
\end{equation}
Due to the inversion symmetry of the continuous beam structure (i.e., if $(\theta_\mathrm{A}^m,\theta_\mathrm{B}^m)^T$ is a solution to Eq.~\eqref{eq24}, then $(\theta_\mathrm{B}^m,\theta_\mathrm{A}^m)^T$ is also a solution to Eq.~\eqref{eq24} with $k$ inverted), the Zak phase $\gamma$ is quantized and interpreted as $\gamma~\mathrm{mod}~2\pi$. Here we note that the usual definition of the Zak phase \citep{Xiao2015exp},
\[ \gamma=\mathrm{i}\int_{-\pi/L}^{\pi/L}dk \langle \varphi \rvert \partial_k \lvert \varphi \rangle \]
where $\varphi(x)$ denotes the Bloch-periodic part of the deflection $\phi(x)$, is in fact equivalent to our expression \eqref{eq7} which uses $\lvert \theta \rangle$ and gives identical results. This is due to the fact that the Zak phase of an inversion-symmetric system depends solely on the \emph{product of parities} of the two modes at the center and the boundary of the Brillouin zone, $k=0$ and $k=\pi/L$ \citep{PhysRevB.83.245132,Lin_2022}.

For odd-numbered bulk bands, $B(\beta l_1)A(\beta l_2)+B(\beta l_2)A(\beta l_1)$ is smaller than zero across the whole range of the band (which is straightforward from the analysis of $\left[\frac{B(\beta l_1)}{A(\beta l_1)}+\frac{B(\beta l_2)}{A(\beta l_2)}\right]$ in Section~3); thus
\begin{equation}
	\label{eqsm10}
	\chi(k) = \pi + \arg \left[C(\beta l_1)A(\beta l_2)+C(\beta l_2)A(\beta l_1) \exp{(\mathrm{i}kL)}\right].
\end{equation}
Then, we obtain
\begin{equation}
	\label{eqsm1}
	\chi(0) = \pi + \arg \left[C(\beta l_1)A(\beta l_2)+C(\beta l_2)A(\beta l_1) \right]\big\vert_{\beta=\beta_b^{(n)}(k=0)},
\end{equation}
where ${\beta=\beta_b^{(n)}(k=0)}$ denotes the frequency of the $n$-th bulk band when $k=0$, and
\begin{equation}
	\label{eqsm2}
	\chi\left(\frac{\pi}{L}\right)=\pi+\arg \left[C(\beta l_1)A(\beta l_2)-C(\beta l_2)A(\beta l_1) \right]\big\vert_{\beta=\beta_b^{(n)}(k=\frac{\pi}{L})},
\end{equation}
where ${\beta=\beta_b^{(n)}(k=\pi/L)}$ denotes the frequency of the $n$-th bulk band when $k=\pi/L$.

For even-numbered bulk bands, $B(\beta l_1)A(\beta l_2)+B(\beta l_2)A(\beta l_1) > 0$; thus
\begin{equation}
	\label{eqsm11}
	\chi(k)=\arg \left[C(\beta l_1)A(\beta l_2)+C(\beta l_2)A(\beta l_1) \exp{(\mathrm{i}kL)}\right].
\end{equation}
Therefore, the expression of the Zak phase for odd- and even-numbered bands is unanimously Eq.~\eqref{eqsm1} minus Eq.~\eqref{eqsm2}, that is, the Zak phases of all frequency bands are obtained as
\[ \gamma^{(n)} = \arg \left[C(\beta l_1)A(\beta l_2)+C(\beta l_2)A(\beta l_1) \right]\big\vert_{\beta=\beta_b^{(n)}(k=0)} - \arg \left[C(\beta l_1)A(\beta l_2)-C(\beta l_2)A(\beta l_1) \right]\big\vert_{\beta=\beta_b^{(n)}(k=\frac{\pi}{L})} . \]

However, the existence of the topological edge states is determined by Eqs.~\eqref{eq20} and \eqref{eq21}, which is equivalent to evaluating
\[ \mu^{(n)}=\arg \left[C(\beta l_1)A(\beta l_2)+C(\beta l_2)A(\beta l_1) \right]\big\vert_{\beta=\beta_t^{(n)}} - \arg \left[C(\beta l_1)A(\beta l_2)-C(\beta l_2)A(\beta l_1) \right]\big\vert_{\beta=\beta_t^{(n)}} \]
at $\beta=\beta_t^{(n)}$; here the result of $\pi$ corresponds to Eq.~\eqref{eq20} with the existent edge states, and $0$ corresponds to Eq.~\eqref{eq21} without edge states. Thus, the existence of the edge states is consistent with the nontrivial Zak phase, if function $[C(\beta l_1)A(\beta l_2)+C(\beta l_2)A(\beta l_1) ]$ has the same sign at $\beta_t^{(n)}$ and $\beta_b^{(n)}(k=0)$, and also function $[C(\beta l_1)A(\beta l_2)-C(\beta l_2)A(\beta l_1) ]$ has the same sign at $\beta_t^{(n)}$ and $\beta_b^{(n)}(k=\pi/L)$; otherwise not. We proceed to give a detailed analysis below with an example structure with $l_1=29~\mathrm{mm}$, and $l_2=50~\mathrm{mm}$, whose eigenvalue spectrum is shown in Fig.~5.

First, from Eq.~\eqref{eq24}, the expressions of the eigenvalues for $H_{\mathrm{Bloch}}^\mathrm{beam}$ are
\begin{align}
	\label{eqsm5}
	&\frac{B(\beta l_1)}{A(\beta l_1)}+\frac{B(\beta l_2)}{A(\beta l_2)}\pm\left[ \frac{C(\beta l_1)}{A(\beta l_1)}+\frac{C(\beta l_2)}{A(\beta l_2)}\right]\qquad (\text{for }kL=0),\\
	\label{eqsm6}
	&\frac{B(\beta l_1)}{A(\beta l_1)}+\frac{B(\beta l_2)}{A(\beta l_2)}\pm\left[ \frac{C(\beta l_1)}{A(\beta l_1)}-\frac{C(\beta l_2)}{A(\beta l_2)}\right]\qquad (\text{for }kL=\pi),
\end{align}
which are represented by orange and blue solid curves in Fig.~5, respectively.

\begin{figure}[tb!]
	\centering
	\includegraphics[width=\linewidth]{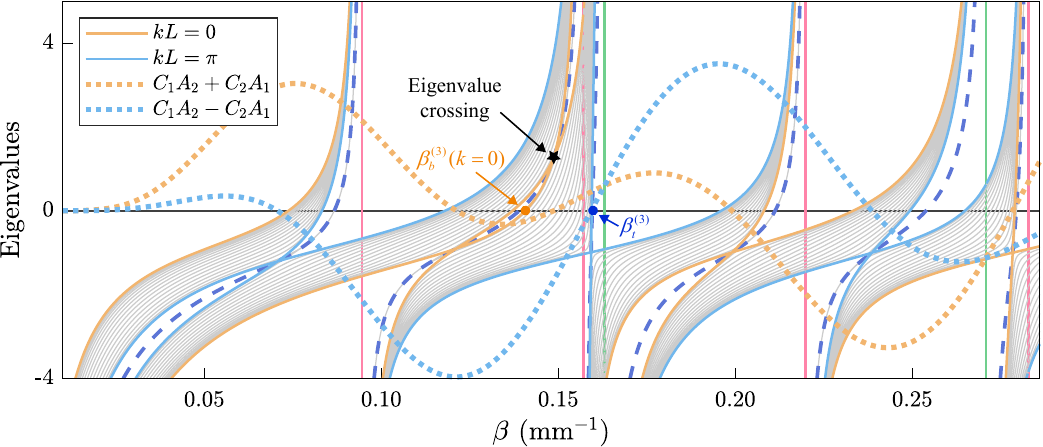}
	\caption{Eigenvalue spectrum of a periodic continuous beam structure with $l_1=29~\mathrm{mm}$, and $l_2=50~\mathrm{mm}$, where an ``eigenvalue crossing'' occurs at a positive eigenvalue (highlighted with a black star); accordingly, the function $[C(\beta l_1)A(\beta l_2)+C(\beta l_2)A(\beta l_1)]$ has opposite signs at $\beta_t^{(3)}$ and $\beta_b^{(3)}(k=0)$. The Zak phase of the third bulk band between frequencies $(\beta_t^{(2)},\beta_t^{(3)})$ is $\pi$, but no topological states exist at $\beta_t^{(3)}$ in the third bandgap. For better visual effects, the functions $[C(\beta l_1)A(\beta l_2) \pm C(\beta l_2)A(\beta l_1)]$ are scaled (by a factor $\exp(-\beta L)$) to be kept inside the plot window.} \label{fig:4}
\end{figure}
Second, we prove that the expressions of the eigenvalues in Eqs.~\eqref{eqsm5} and \eqref{eqsm6} increase monotonically with $\beta$ (i.e., the orange and blue curves always have positive slopes). It suffices to prove that functions $\left[\frac{B}{A}(\beta l)+\frac{C}{A}(\beta l)\right]$ and $\left[\frac{B}{A}(\beta l)-\frac{C}{A}(\beta l)\right]$ always have positive derivatives with respect to its argument $\beta l$. We have
\begin{align}
	\label{eqsm12}
	\left[\frac{B(\beta l)}{A(\beta l)}+\frac{C(\beta l)}{A(\beta l)}\right]'&=\frac{(\cosh{\beta l}+\cos{\beta l}+2)(\cosh{\beta l}-\cos{\beta l}-\sinh{\beta l}\sin{\beta l})}{A^2(\beta l)},\\
	\label{eqsm13}
	\left[\frac{B(\beta l)}{A(\beta l)}-\frac{C(\beta l)}{A(\beta l)}\right]'&=\frac{(\cosh{\beta l}+\cos{\beta l}-2)(\cosh{\beta l}-\cos{\beta l}+\sinh{\beta l}\sin{\beta l})}{A^2(\beta l)}.
\end{align}
Because $\cosh{\beta l}+\cos{\beta l}\pm 2 >0$ for all $\beta>0$, we only need to prove that
\begin{align}
	\label{eqsm14}
	h_1(x) &\equiv \cosh{\beta l}-\cos{\beta l}-\sinh{\beta l}\sin{\beta l}, \\
	\label{eqsm15}
	h_2(x) &\equiv \cosh{\beta l}-\cos{\beta l}+\sinh{\beta l}\sin{\beta l}
\end{align}
are non-negative. We make the substitutions
\[ t\equiv \tan\frac{\beta l}{2},\quad u\equiv \tanh\frac{\beta l}{2}, \]
so that
\[ \sin{\beta l} = \frac{2t}{1+t^2}, \quad \cos{\beta l} = \frac{1-t^2}{1+t^2}, \quad \sinh{\beta l} = \frac{2u}{1-u^2}, \quad \cosh{\beta l} = \frac{1+u^2}{1-u^2}; \]
thus, we obtain
\begin{align}
	\label{eqsm16}
	h_1(x) &= \frac{2(u-t)^2}{(1+t^2)(1-u^2)} \ge 0, \\
	\label{eqsm17}
	h_2(x) &= \frac{2(u+t)^2}{(1+t^2)(1-u^2)} \ge 0,
\end{align}
which are non-negative, and equal to zero only at discrete points $u\pm t=0$ (which are in fact the zeros of $A(\beta l) = \frac{2(t^2-u^2)}{(1+t^2)(1-u^2)}$).

Third, the ``eigenvalue crossings'' (i.e., eigenvalues becoming degenerate) for $kL=0$ -- represented by intersection points of the two orange curves -- correspond to the zeros of $[C(\beta l_1)A(\beta l_2)+C(\beta l_2)A(\beta l_1)]$ (orange dashed line), and those for $kL=\pi$ -- represented by intersection points of the two blue curves -- correspond to the zeros of $[C(\beta l_1)A(\beta l_2)-C(\beta l_2)A(\beta l_1)]$ (blue dashed line).
\begin{figure}[tb!]
	\centering
	\includegraphics[width=0.8\linewidth]{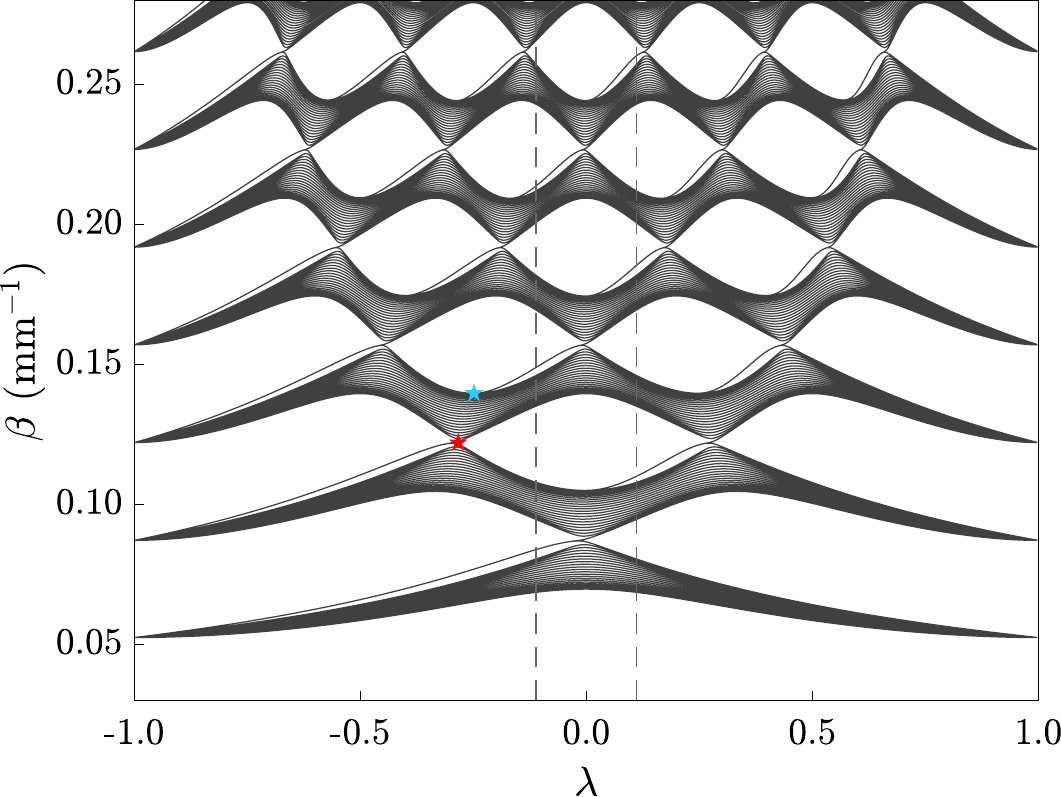}
	\caption{Frequency spectrum of the continuous beam structure with respect to the geometric parameter $\lambda$, where $\lambda$ is defined such that the lengths in the structure are $l_1=(1+\lambda) l_\mathrm{ave}$ and $l_2=(1-\lambda) l_\mathrm{ave}$ (here we fix $l_\mathrm{ave} = 45~\mathrm{mm}$). Two dashed lines $\lambda=\pm1/9$ correspond to the frequency spectra of the two structures in Fig.~1(b),(c). The red and blue stars represent two types of points across which the topological edge states emerge or vanish.} \label{fig:5}
\end{figure}

Finally, from the above arguments, we rigorously conclude that when eigenvalue crossing occurs \emph{only at negative values} $y<0$, the existence of the topological edge states in the $n$-th bandgap is directly determined by the Zak phase of the $n$-th bulk band; however, if an eigenvalue crossing occurs \emph{at a positive value} $y>0$ in the interval $(\beta_t^{(n-1)},\beta_0^{(n-1)})$, the existence of the topological edge state in the $n$-th bandgap is not consistent with the Zak phase, but instead determined by the topological index $\mu^{(n)}$ defined by Eq.~\eqref{eqindex}:
\[
\mu^{(n)}=\gamma^{(n)}+\pi\cdot l^{(n)} \mod 2\pi,
\]
where $\gamma^{(n)}$ denotes the Zak phase of the $n$-th bulk band, and $l^{(n)}$ is the number of the eigenvalue crossings occurring at the positive eigenvalues within the frequency range of the $n$-th bulk band.

To summarize, we use Fig.~6 to show the topological phase transitions of continuous beam structures and the relation between the topological phases and the existence of the edge states. The red star in Fig.~6 represents the band-crossing point (i.e., the topological phase transition point $\Big \lvert\frac{C(\beta_t l_1)}{A(\beta_t l_1)}\Big \rvert=\Big \lvert\frac{C(\beta_t l_2)}{A(\beta_t l_2)}\Big \rvert$). The blue star represents the \emph{point of the coincidental zeros of $A(\beta l_1)$ and $A(\beta l_2)$}. The existence of the topological edge states (black thin lines separated from the bands) will change at these two types of points, as clearly indicated in Fig.~6.
When the geometric parameter $\lambda$ of the structure passes through the band-crossing points, the Zak phases of the adjacent frequency bands are altered. The phenomenon of an ``eigenvalue crossing'' at a positive eigenvalue will occur for $\lambda$ within the narrow range between the red star ($\lambda=-0.2831$) and the blue star ($\lambda=-0.2482$), for the third bandgap in this case. At the band-crossing point (red star), the position of the ``eigenvalue crossing'' moves to the $y=0$ axis; across the point where the zeros of $A(\beta l_1)$ and $A(\beta l_2)$ coincide (blue star), the eigenvalue crossing point quickly moves to positive infinity and vanish. Therefore, the existence of the topological edge states is jointly determined by the Zak phase and the eigenvalue crossing, consistent with our criterion Eq.~\eqref{eqindex}.

\section{Topological dynamics of general continuum lattice grid structures}
The theoretical framework we built above for continuous beams is powerful in that it is directly applicable to reveal the topological dynamics of general continuum lattice grid structures, such as bridge-like frames (Fig.~7), square frames (Fig.~8(a),(b)), kagome frames (Fig.~8(c),(d)), etc., and even homogeneous plates. We use the above framework to identify the frequencies of their topological edge or corner states, give a criterion for phase transitions, and demonstrate the topological states with analytical results.
\subsection{Theoretical framework for continuum lattice grid structures}
For a general planar continuum lattice grid structure consisting of intersecting and joined beams, such as bridge-like, square, and kagome frames (Figs.~7 and 8), and continuous beams on elastic springs (Fig.~9(a),(b)), the theoretical analysis works in an analogous manner: one takes all the rotations of the joints as basic variables (e.g., represented by $\theta_i$), and the dynamic matrix can be expressed in a straightforward manner by the following simple rules:
\begin{itemize}
	\item The $(i,j)$ nondiagonal entries are $-\frac{C}{A}(\beta l_{ij})$ (where $l_{ij}$ denotes the length of a beam between two joints $i$ and $j$, if present);
	\item the $(i,i)$ diagonal entries are $\sum_j \frac{B}{A}(\beta l_{ij})$ (where all beam segments around joint $i$ are taken into account); and
	\item the remaining entries are left as zero.
	\item If torsional springs $k_t$ are present at joints, add $-\frac{k_t}{\beta EI}$ to the corresponding diagonal entries;
	\item if Winkler elastic foundations $k_f$ are present at beam segments (e.g., segment between joints $i$ and $j$), replace $\frac{B}{A}(\beta l_{ij})$ by $\frac{\tilde{\beta}}{\beta} \cdot \frac{B}{A}(\tilde{\beta} l_{ij})$ and $-\frac{C}{A}(\beta l_{ij})$ by $-\frac{\tilde{\beta}}{\beta} \cdot \frac{C}{A}(\tilde{\beta} l_{ij})$ for those beam segments, where $\tilde{\beta}$ is defined by $\tilde{\beta}^4 = \beta^4-\frac{k_f}{EI}$.
\end{itemize}
Such a procedure reduces the complex infinite-degree-of-freedom problem into a small matrix (after performing Bloch analysis) analogous to the Bloch Hamiltonian in finite-degree-of-freedom topological models. Moreover, all matrix elements are analytical expressions, directly obtained from the governing differential equations for the beam segments. This renders our formulation both analytical and precise. Note that we require all joints to be constrained so as to admit no translational displacements.
\subsection{Topological edge states of bridge-like frames}
We first study a bridge-like frame structure (Fig.~7). The horizontal and vertical beam segments are uniform (though our theory is  applicable to the cases where the segments have different thicknesses) and have lengths $l_{1,2}$ and $l_0$, respectively. All boundaries for the outermost ends are clamped. The vector $\lvert\theta\rangle = (\theta_1, \theta_2, \dots, \theta_{2N-1}, \theta_{2N})^T$ comprises the rotations of the $2N$ intersection joints of the horizontal beam with all $2N$ vertical beams (note that all clamped ends, which do not admit rotations, are excluded from $\lvert\theta\rangle$). Then the dynamic matrix $H^{\text{frame}}$ for the governing equation $H^{\text{frame}} \lvert\theta\rangle = 0$ of the structure is
\[ H^{\text{frame}} =
\begin{bmatrix}
	\sum\limits_{i=0}^2 \frac{B(\beta l_i)}{A(\beta l_i)} & -\frac{C(\beta l_1)}{A(\beta l_1)} & 0 & 0 & \cdots & 0 & 0 & 0 \\
	-\frac{C(\beta l_1)}{A(\beta l_1)} & \sum\limits_{i=0}^2 \frac{B(\beta l_i)}{A(\beta l_i)} & -\frac{C(\beta l_2)}{A(\beta l_2)} & 0 & \cdots & 0 & 0 & 0 \\
	0 & -\frac{C(\beta l_2)}{A(\beta l_2)} & \sum\limits_{i=0}^2 \frac{B(\beta l_i)}{A(\beta l_i)} & -\frac{C(\beta l_1)}{A(\beta l_1)} & \cdots & 0 & 0 & 0 \\
	0 & 0 & -\frac{C(\beta l_1)}{A(\beta l_1)} & \sum\limits_{i=0}^2 \frac{B(\beta l_i)}{A(\beta l_i)} & \cdots & 0 & 0 & 0 \\
	\vdots & \vdots & \vdots & \vdots & & \vdots & \vdots & \vdots \\
	0 & 0 & 0 & 0 & \cdots & \sum\limits_{i=0}^2 \frac{B(\beta l_i)}{A(\beta l_i)} & -\frac{C(\beta l_2)}{A(\beta l_2)} & 0 \\
	0 & 0 & 0 & 0 & \cdots & -\frac{C(\beta l_2)}{A(\beta l_2)} & \sum\limits_{i=0}^2 \frac{B(\beta l_i)}{A(\beta l_i)} & -\frac{C(\beta l_1)}{A(\beta l_1)} \\
	0 & 0 & 0 & 0 & \cdots & 0 & -\frac{C(\beta l_1)}{A(\beta l_1)} & \sum\limits_{i=0}^2 \frac{B(\beta l_i)}{A(\beta l_i)}
\end{bmatrix},
\]
where the functions $A$, $B$, and $C$ are defined in Eqs.~\eqref{eqa2}--\eqref{eqa4}, and $l_{0,1,2}$ denote the lengths of the beam segments (see Fig.~7(a)); the corresponding Bloch matrix is then
\[ H^{\text{frame}}_{\text{Bloch}} =
\begin{bmatrix}
	\displaystyle \sum\limits_{i=0}^2 \frac{B(\beta l_i)}{A(\beta l_i)} & \displaystyle -\frac{C(\beta l_1)}{A(\beta l_1)} - \frac{C(\beta l_2)}{A(\beta l_2)} \exp{(-\mathrm{i}kL)} \\
	\displaystyle -\frac{C(\beta l_1)}{A(\beta l_1)} - \frac{C(\beta l_2)}{A(\beta l_2)} \exp{(\mathrm{i}kL)} & \displaystyle \sum\limits_{i=0}^2 \frac{B(\beta l_i)}{A(\beta l_i)}
\end{bmatrix}.
\]
\begin{figure}[tb!]
	\centering
	\includegraphics[width=0.8\linewidth]{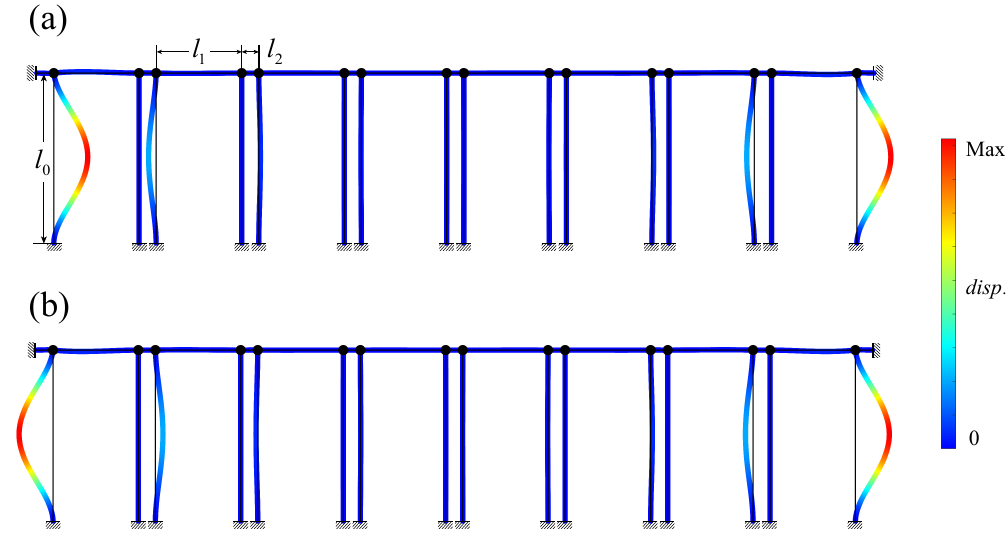}
	\caption{Topological edge states in a bridge-like frame structure with lengths of horizontal beam segments $l_1=50~\mathrm{mm}$, $l_2=10~\mathrm{mm}$, and vertical beams $l_0=100~\mathrm{mm}$. The rotational angles are symmetrically (a) and antisymmetrically (b) distributed, which are localized at the leftmost and rightmost joints. The thin black lines indicate the configuration of the frames before deformation.} \label{fig:6:1}
\end{figure}They have an SSH-chain-like form in the subdiagonals, and also the nice property of identical diagonal elements that facilitates  analysis. Similar to the topological analysis for the continuous beam structures, the topological edge states emerge at $\beta_t$ which satisfies
\begin{equation}
	\frac{B(\beta_t l_0)}{A(\beta_t l_0)} + \frac{B(\beta_t l_1)}{A(\beta_t l_1)} + \frac{B(\beta_t l_2)}{A(\beta_t l_2)} = 0 \quad \text{and} \quad \left \lvert\frac{C(\beta_t l_1)}{A(\beta_t l_1)}\right \rvert < \left \lvert\frac{C(\beta_t l_2)}{A(\beta_t l_2)}\right \rvert.
\end{equation}
We find that the bulk bands are separated from the topological edge states when $\Big \lvert\frac{C(\beta_t l_1)}{A(\beta_t l_1)}\Big \rvert\neq\Big \lvert\frac{C(\beta_t l_2)}{A(\beta_t l_2)}\Big \rvert$, and the topological phase transition points are at $\Big \lvert\frac{C(\beta_t l_1)}{A(\beta_t l_1)}\Big \rvert=\Big \lvert\frac{C(\beta_t l_2)}{A(\beta_t l_2)}\Big \rvert$. As shown in Fig.~7(a),(b), degenerate topological edge states in the first bandgap exist when $l_1>l_2$ and $l_{1,2}\ll l_0$.
\begin{figure}[t!]
	\centering
	\includegraphics[width=0.9\linewidth]{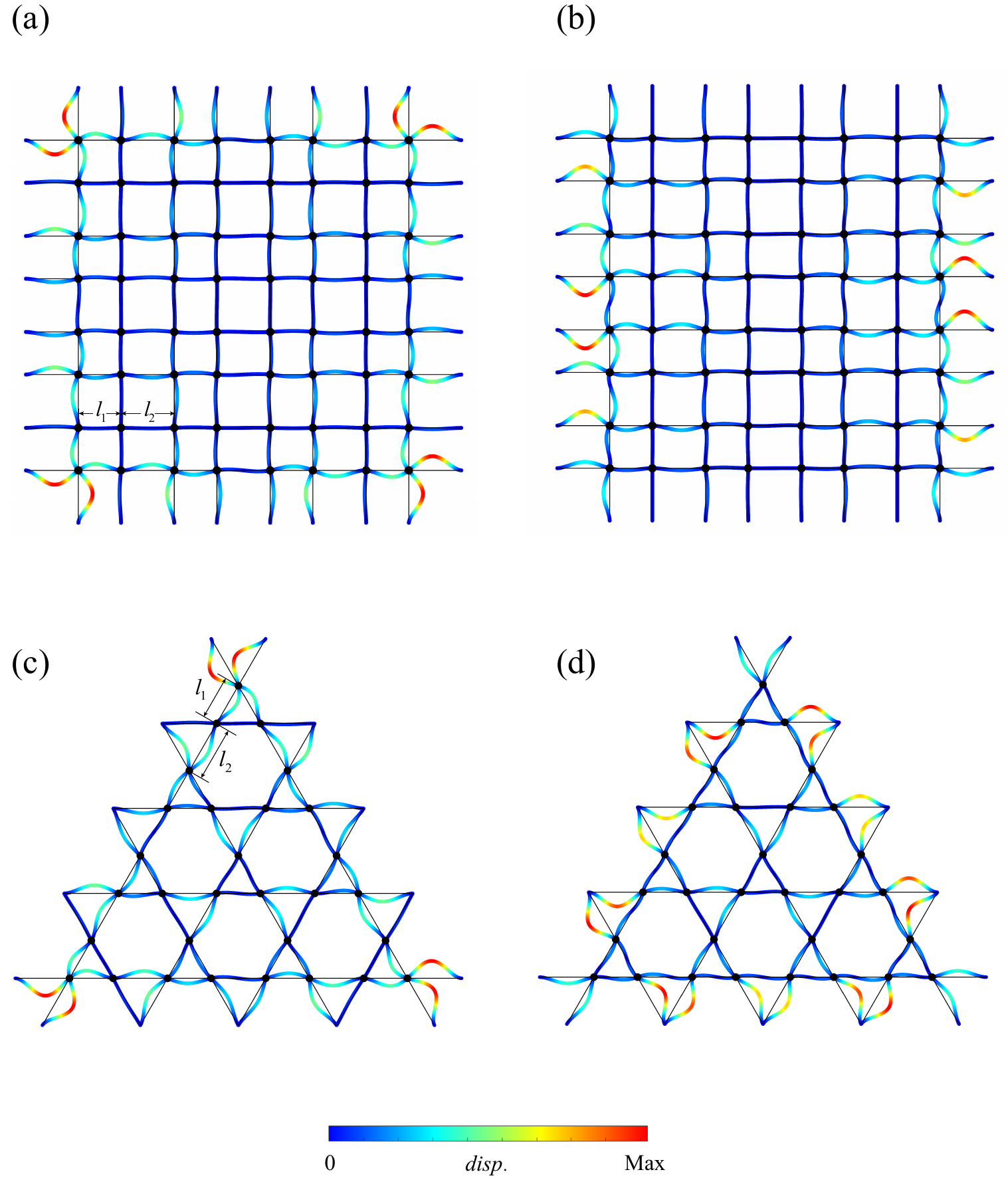}
	\caption{Higher-order topological dynamics of square frame and kagome frames. (a) A topological corner state and (b) a topological edge state in the square frame  with $l_1=40~\mathrm{mm}$, $l_2=50~\mathrm{mm}$. (c) A topological corner state and (d) a topological edge state in the kagome frame with $l_1=40~\mathrm{mm}$, $l_2=50~\mathrm{mm}$.} \label{fig:6:2}
\end{figure}

\subsection{Higher-order topological states of square frames}
A higher-order topological insulator -- the square frame structure with $C_4$ symmetry, as shown in Fig.~8(a),(b), is constructed with alternately arranged beam segments with lengths $l_1$ and $l_2$ in two directions. All the outermost ends are clamped. The Bloch dynamic matrix for the square frame structure is
	\[
	H_{\mathrm{Bloch}}^\mathrm{square}=H_{\mathrm{Bloch}}^\mathrm{beam}(k_x) \otimes I + I \otimes H_{\mathrm{Bloch}}^\mathrm{beam}(k_y),
	\]
	where $k_x$ and $k_y$ are wavenumbers in the orthogonal directions.
	Since the square lattice has chiral symmetry \citep{Floquetsum}, zero-dimensional topological corner states emerge at the frequencies $\beta_t$ where the main-diagonal elements of the dynamic matrix satisfy
	\begin{equation}
		2 \left[\frac{B(\beta l_1)}{A(\beta l_1)}+\frac{B(\beta l_2)}{A(\beta l_2)}\right] = 0
	\end{equation}
	and the non-diagonal elements satisfy
	\begin{equation} \label{eqhigherorder}
		\left \lvert \frac{C(\beta l_1)}{A(\beta l_1)} \right \rvert < \left \lvert\frac{C(\beta l_2)}{A(\beta l_2)} \right \rvert.
	\end{equation}
Also, one-dimensional topological edge states are present at the frequencies satisfying
	\begin{equation}
		2\left\lvert \frac{B(\beta l_1)}{A(\beta l_1)}+\frac{B(\beta l_2)}{A(\beta l_2)} \right\rvert = \left\lvert-\frac{C(\beta l_1)}{A(\beta l_1)} - \frac{C(\beta l_2)}{A(\beta l_2)} \exp{(-\mathrm{i}k_{x,y}L)}\right\rvert ,
	\end{equation}
as long as the condition \eqref{eqhigherorder} also holds. We find that the corner states are separated from the topological edge states when $\Big \lvert\frac{C(\beta_t l_1)}{A(\beta_t l_1)}\Big \rvert\neq\Big \lvert\frac{C(\beta_t l_2)}{A(\beta_t l_2)}\Big \rvert$, and the topological phase transition points are at $\Big \lvert\frac{C(\beta_t l_1)}{A(\beta_t l_1)}\Big \rvert=\Big \lvert\frac{C(\beta_t l_2)}{A(\beta_t l_2)}\Big \rvert$. The topological corner states are localized at the four corners of the square frame, as shown in Fig.~8(a). The topological edge states are localized at the edges as shown in Fig.~8(b).
\subsection{Higher-order topological states of kagome frames}
A kagome frame structure with $C_3$ symmetry, as shown in Fig.~8(c),(d), is constructed with alternately arranged beam segments with lengths $l_1$ and $l_2$. All the outermost ends are clamped. The Bloch dynamic matrix  is
	\[ H^{\text{kagome}}_{\text{Bloch}} =
	\begin{bmatrix}
		2\sum\limits_{i=1}^2 \frac{B(\beta l_i)}{A(\beta l_i)} & -\frac{C(\beta l_1)}{A(\beta l_1)} - \frac{C(\beta l_2)}{A(\beta l_2)} e^{-\mathrm{i}k_xL} &
		-\frac{C(\beta l_1)}{A(\beta l_1)} - \frac{C(\beta l_2)}{A(\beta l_2)} e^{\mathrm{i}(-\frac{1}{2}k_x-\frac{\sqrt{3}}{2}k_y)L}
		\\
		-\frac{C(\beta l_1)}{A(\beta l_1)} - \frac{C(\beta l_2)}{A(\beta l_2)} e^{\mathrm{i}k_xL} & 2\sum\limits_{i=1}^2 \frac{B(\beta l_i)}{A(\beta l_i)}
		&  -\frac{C(\beta l_1)}{A(\beta l_1)} - \frac{C(\beta l_2)}{A(\beta l_2)} e^{\mathrm{i}(\frac{1}{2}k_x-\frac{\sqrt{3}}{2}k_y)L} \\
		-\frac{C(\beta l_1)}{A(\beta l_1)} - \frac{C(\beta l_2)}{A(\beta l_2)} e^{\mathrm{i}(\frac{1}{2}k_x+\frac{\sqrt{3}}{2}k_y)L} &
		-\frac{C(\beta l_1)}{A(\beta l_1)} - \frac{C(\beta l_2)}{A(\beta l_2)} e^{\mathrm{i}(-\frac{1}{2}k_x+\frac{\sqrt{3}}{2}k_y)L} &  2\sum\limits_{i=1}^2 \frac{B(\beta l_i)}{A(\beta l_i)}
	\end{bmatrix}.
	\]
Since the kagome lattice has generalized chiral symmetry \citep{kagome}, topological corner states exist at frequencies $\beta_t$ where main-diagonal elements of the dynamic matrix $2\left[\frac{B(\beta_t l_1)}{A(\beta_t l_1)}+\frac{B(\beta_t l_2)}{A(\beta_t l_2)}\right]=0$ and also $\Big \lvert\frac{C(\beta_t l_1)}{A(\beta_t l_1)}\Big \rvert<\Big \lvert\frac{C(\beta_t l_2)}{A(\beta_t l_2)}\Big \rvert$. Topological corner states and edge states are illustrated in Fig.~8(c),(d).

\subsection{Topological edge states of continuous beam structures on elastic springs}
\begin{figure}[tb]
	\centering
	\includegraphics[width=0.8\linewidth]{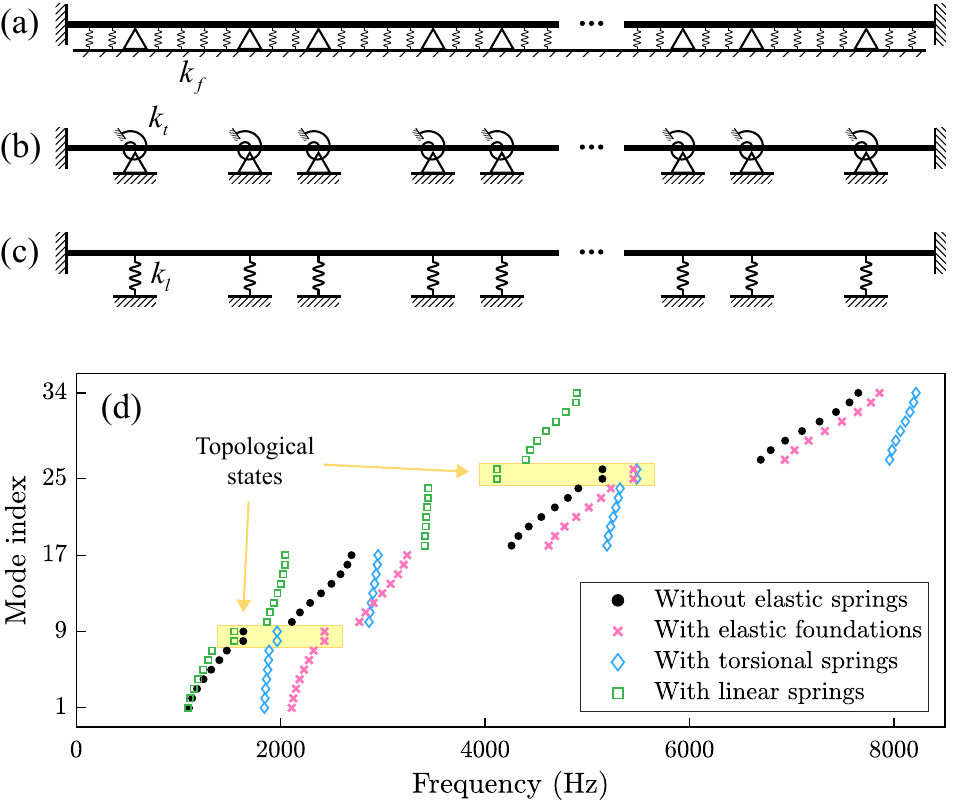}
	\caption{(a)--(c) Continuous beams on the Winkler elastic foundation, torsional springs, and linear springs. (d) Frequency spectra of the three structures. The eigenfrequencies of the topological states are shifted from those of the original structure when springs are added. (Parameters for the plot: $k_f= 375EI/l_1^4$, $k_t=30EI/l_1$, and $k_l=750EI/l_1^3$, with $l_1=40~\mathrm{mm}$ and $l_2=50~\mathrm{mm}$.)} \label{fig:9}
\end{figure}
The theoretical method is also extended to variants of continuous beam structures by introducing elastic springs, so that we can achieve precise \emph{modulations} of the frequencies of the topological states. If a continuous beam is supported on the \emph{Winkler elastic foundation} with spring constant $k_f$ as shown in Fig.~9(a), the relation between the eigenfrequencies of the beam structures with and without elastic foundations is $\omega_f^2=\omega^2+k_f/m$. If \emph{torsional springs} $k_t$ are present at the joints as shown in Fig.~9(b), $\beta_t^{(n)}$ of the topological states becomes larger compared to $\beta$ of the continuous beam without the springs; however, it does not exceed the upper bound $\beta_0^{(n)}$ of the interval $(\beta_0^{(n-1)},\beta_0^{(n)})$. If a continuous beam is supported on discrete \emph{linear springs} $k_l$, as shown in Fig.~9(c), the linear springs are regarded as a disturbance to the original structure (analogous to the effect of next-nearest-neighbor hoppings for extended SSH chains). In this case, the degree of localization of the topological states decreases with the decrease of $k_l$.
The frequency spectra containing the topological states in the two lowest-order mode shapes for the three types of supported beams are shown in Fig.~9(d).

\subsection{Topological edge states of homogeneous plates}
For a homogeneous plate where the edges $y=0$ and $y=b$ are simply supported, the edges $x=0$ and $x=N(l_1+l_2)+l_2$ are clamped, and simple supports are set to separate it into lengths $l_2$, $l_1$, $l_2$, \dots, $l_1$, $l_2$ along the $x$-direction (Fig.~10), it has decoupled deflections in two orthogonal directions, with nontrivial topological properties along the $x$-direction. Each section of the plate perpendicular to the $y$-axis has a mode profile localized near the edges $x=0$ and $x=N(l_1+l_2)+l_2$. The two localized topological modes are shown in Fig.~10(a),(b).
\begin{figure}[tb!]
	\centering
	\includegraphics[width=0.7\linewidth]{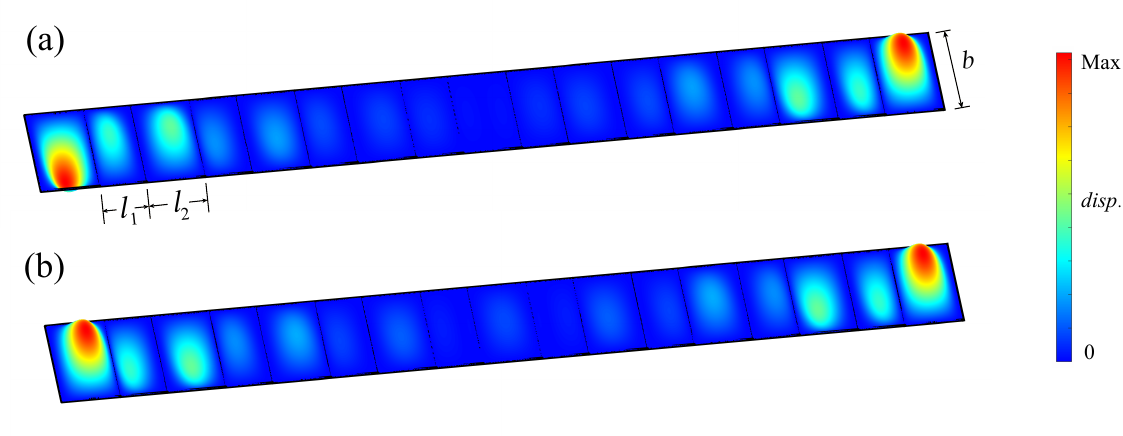}
\caption{Topological edge modes of a supported plate with interval lengths $l_1=40~\mathrm{mm}$, $l_2=50~\mathrm{mm}$, and width $b=100~\mathrm{mm}$. The rotational angles are symmetrically (a) and antisymmetrically (b) distributed, which are localized at the leftmost and rightmost joints. } \label{fig:6:10}
\end{figure}

\section{Conclusions}
\label{sec5}
In this work, we establish a theoretical framework of the topological dynamics of continuum lattice grid structures, and rigorously identify the infinitely many topological edge states within the bandgaps, and introduce a topological index related to the bulk geometric phases that determines the existence of edge states. We clearly reveal the topological phase transitions and the relation between the existence of edge states and the topological phases for bulk bands. The theoretical approach is applied to reveal the topological dynamics of several typical continuum lattice grid structures including bridge-like frames, square frames, kagome frames, and continuous beams supported on elastic foundations, torsional springs, and linear springs, as well as homogeneous plates, which demonstrates the universality and applicability of our theory.  Accurate analytical expressions of the frequencies of topological edge and corner states are obtained, applicable to any band from low- to high-frequency. The continuum lattice grid structures serve as excellent platforms for exploring various kinds of topological phases and demonstrating the topologically protected states at multiple frequencies, due to their simple configurations and the concise theoretical framework. Continuous beam and frame structures widely exist in buildings, bridges, railways, ships, advanced microstructural materials etc., and thus the topological dynamics of these continuum lattice grid structures has significant implications in safety assessment, structural health monitoring, and energy harvesting, etc.

\section*{CRediT authorship contribution statement}
\textbf{Yimeng Sun:} Conceptualization, Methodology, Software, Formal analysis, Validation, Writing – original draft, Visualization. \textbf{Jiacheng Xing}: Methodology, Software, Formal analysis, Validation, Writing – original draft, Visualization. \textbf{Li-Hua Shao:} Resources, Formal analysis, Validation, Writing – review \& editing, Funding acquisition. \textbf{Jianxiang Wang:} Conceptualization, Methodology, Formal analysis, Validation, Resources, Supervision, Writing – review \& editing, Funding acquisition.
\section*{Declaration of Competing Interest}
The authors declare that they have no known competing financial interests or personal relationships that could have appeared to influence the work reported in this paper.
\section*{Data availability}
Data will be made available on request.
\section*{Acknowledgements}
Y. S., J. X. and J. W. thank the National Natural Science Foundation of China (Grants No. 11991033), and L.-H. S. thanks the Beijing Natural Science Foundation (Grant No. JQ21001), for support of this work.





\bibliographystyle{elsarticle-harv}
\bibliography{sunyimeng}


\end{document}